\documentclass[twocolumn,txfonts,showpacs,nofootinbib,superscriptaddress,preprintnumbers,amsmath,amssymb,epsfig,color]{aa}

\usepackage{natbib}
\bibpunct{(}{)}{;}{a}{}{,} % to follow the A&A style...

\usepackage[usenames]{color}
\usepackage{graphicx}% Include figure files

\usepackage{units, subfigure,amssymb,amsmath}
\usepackage[latin1]{inputenc}
\def\vectheta{\vec{\theta}}

\begin{document}
\input epsf
\title{A Markov Chain Monte Carlo technique to sample transport and source
parameters of Galactic cosmic rays}
\subtitle{II. Results for the diffusion model combining B/C and radioactive nuclei}
\author{
       A. Putze\inst{1,2}
	\and L. Derome\inst{2}
	\and D. Maurin\inst{3,4,5}
} 
\offprints{Antje Putze, {\tt antje@fysik.su.se}}

\institute{
   The Oskar Klein Centre for Cosmoparticle Physics,
   Department of Physics, Stockholm University,
   AlbaNova, SE-10691 Stockholm, Sweden
   \and Laboratoire de Physique Subatomique et de
	Cosmologie ({\sc lpsc}),
	Universit\'e Joseph Fourier Grenoble 1, CNRS/IN2P3, Institut Polytechnique de Grenoble,
	53 avenue des Martyrs,
	Grenoble, 38026, France
   \and Laboratoire de Physique Nucl\'eaire et des Hautes
	Energies ({\sc lpnhe}),
   Universit\'es Paris VI et Paris VII, CNRS/IN2P3,
	Tour 33, Jussieu, Paris,
	75005, France
   \and
   Dept. of Physics and Astronomy, University of Leicester,
   Leicester, LE17RH, UK
   \and
   Institut d'Astrophysique de Paris ({\sc iap}), UMR7095 CNRS,
   Universit\'e Pierre et Marie Curie, 98 bis bd Arago,
   75014 Paris, France
}

\date{Received / Accepted}% It is always \today, today,

\abstract
%Context
{Ongoing measurements of the cosmic radiation (nuclear, electronic, and $\gamma$-ray)
are providing additional insight into cosmic-ray physics. A comprehensive picture of these data
relies on an accurate determination of the transport and source parameters of propagation
models.
}
%Text of aims
{A Markov Chain Monte Carlo method is used to obtain these parameters in a diffusion model.
By measuring the B/C ratio and radioactive cosmic-ray clocks, we calculate their
probability density functions, placing special emphasis on the halo size $L$ of the Galaxy
and the local underdense bubble of size $r_h$. We also derive the mean, best-fit model parameters
and 68\% confidence level for the various parameters, and the envelopes of other quantities.
} %Text of methods
{The analysis relies on the USINE code for propagation and on a Markov Chain Monte Carlo
technique previously developed by ourselves for the parameter determination.}
%Text of results
{The B/C analysis leads to a most probable diffusion slope $\delta \!=\!
0.86^{+0.04}_{-0.04}$ for diffusion, convection, and reacceleration, or $\delta =
0.234^{+0.006}_{-0.005}$ for diffusion and reacceleration. As found in previous
studies, the B/C best-fit model favours the first configuration, hence pointing to a
high value for $\delta$. These results do not depend on $L$, and we provide simple
functions to rescale the value of the transport parameters to any $L$. A combined fit on
B/C and the isotopic ratios ($^{10}$Be/$^9$Be, $^{26}$Al/$^{27}$Al, $^{36}$Cl/Cl) leads
to $L=8^{+8}_{-7}$~kpc and $r_{h}= 120^{+20}_{-20}$~pc for the best-fit model. This
value for $r_h$ is consistent with direct measurements of the local interstallar medium.
For the model with diffusion and reacceleration, $L=4^{+1}_{-1}$~kpc and $r_{h}=3^{+70}_{-3}$~pc (consistent with zero).
We vary $\delta$, because its value is still disputed. For the model with Galactic winds,
we find that between $\delta=0.2$ and 0.9, $L$ varies from ${\cal O}(0)$ to ${\cal O}(2)$
if $r_h$ is forced to be 0,  but it otherwise varies from ${\cal O}(0)$ to  ${\cal O}(1)$ (with
$r_h\sim100$~pc for all $\delta\gtrsim0.3$). The results from the elemental ratios Be/B,
Al/Mg, and Cl/Ar do not allow independent checks of this picture because these data are
not precise enough.
}
% Conclusions
{We showed the potential and usefulness of the Markov Chain Monte Carlo technique in the analysis of cosmic-ray
measurements in diffusion models. The size of the diffusive halo depends crucially on the
value of the diffusion slope $\delta$, and also on the presence/absence of the local
underdensity damping effect on radioactive nuclei. More precise data from ongoing
experiments are expected to clarify this issue.
}


\titlerunning{An MCMC technique to sample transport and source parameters of Galactic cosmic rays. II.}
\maketitle

%\tableofcontents

%%%%%%%%%%%%%%%%%%%%%%%%%%%%%%%%%%%%%%%%%%%%%%%%%%%%%%%%%%%%%%%%%%%%%%
%%%%%%%%%%%%%%%%%%%%%%%%%%%%%%%%%%%%%%%%%%%%%%%%%%%%%%%%%%%%%%%%%%%%%%
\section{Introduction}
Almost a century after the discovery of cosmic radiation, the number of precision
instruments devoted to Galactic cosmic ray (GCR) measurements in the GeV-TeV energy range
is unprecedented. 
The GeV $\gamma$-ray diffuse emission is being measured by the {\sc Fermi} satellite
\citep{2009arXiv0912.0973T}, while the
TeV diffuse emission is within reach of ground arrays of Cerenkov Telescopes (e.g.,
{\sc Hess}, \citealt{2006Natur.439..695A}; {\sc Milagro}, \citealt{2008ApJ...688.1078A}).
The high-energy spectrum of electrons and positrons uncovered some surprising and still
debated features ({\sc Atic}, \citealt{2008Natur.456..362C}; {\sc Fermi}, \citealt{2009arXiv0905.0025F};
{\sc Hess}, \citealt{2008PhRvL.101z1104A,2009arXiv0905.0105H}; 
{\sc Pamela}, \citealt{2009Natur.458..607A}; {\sc Ppp-bets}, \citealt{2008arXiv0809.0760T}).
For nuclei, many experiments (satellites and balloon-borne) have acquired data, that remain to be
published ({\sc Cream}, \citealt{2008APh....30..133A}; {\sc Tracer}, \citealt{2008ApJ...678..262A};
{\sc Atic}, \citealt{2008ICRC....2....3P}; {\sc Pamela}). 
Anti-protons are also being measured ({\sc Pamela}, \citealt{2009PhRvL.102e1101A}) and are targets
for future satellite and balloon experiments ({\sc Ams-02}, {\sc Bess}-Polar). Anti-deuteron detection
should be achieved in a few years ({\sc Ams-02}, \citealt{2008ICRC....4..765C};
{\sc Gaps}, \citealt{2008AdSpR..41.2056F}).
A complementary view of cosmic-ray propagation is given by anisotropy measurements from
ground experiments of high energy (e.g., the Tibet Air Shower
Arrays, \citealt{2006Sci...314..439A}; Super-Kamiokande-I
detector, \citealt{2007PhRvD..75f2003G}; {\sc Eas-top}, \citealt{2009ApJ...692L.130A}). 
This multi-messenger and multi-energy picture will soon be completed: neutrino detectors are still in
development (e.g., {\sc Icecube}, {\sc Km3n}e{\sc t}), but identifying the sources of the GCRs should
be within reach a few years after data collection \citep{2008PhRvD..78f3004H}.

All these measurements are probes to understanding and uncovering the sources of cosmic rays, the
mechanisms of propagation, and the interaction of CRs with the gas and the radiation field of 
the Galaxy \citep{2007ARNPS..57..285S}. It is important to determine the propagation parameters,
because their value can be compared to theoretical predictions for the transport in turbulent
magnetic fields (e.g.,
\citealt{2002PhRvD..65b3002C,2006ApJ...642..902P,2007ApJ...663.1049M,2008ApJ...672..642T},
\citealt{2008ApJ...673..942Y} and references therein), or related to the source spectra
predicted in acceleration models (e.g.,
\citealt{2006A&A...453..193M,2007Natur.449..576U,2008Natur.453...48P,2008MNRAS.386..509R},
\citealt{2008ARA&A..46...89R} and references therein). The transport and source parameters are
also related to Galactic astrophysics (e.g., nuclear abundances and stellar
nucleosynthesis|\citealt{1990PhR...191..353S,1997SSRv...81..107W}), and to dark matter
indirect detection \citep[e.g.,][]{2004PhRvD..69f3501D,2008PhRvD..77f3527D}.

In the first paper of this series (\citealt{2009A&A...497..991P}, hereafter Paper~I), we implemented
a Markov Chain Monte Carlo (MCMC) to estimate the probability density function (PDF) of the
transport and source parameters. This allowed us to constrain these parameters with a sound statistical
method, to assess the goodness of fit of the models, and as a by-product, to provide 68\% and
95\% confidence level (CL) envelopes for any quantity we are interested in (e.g., B/C ratio,
anti-proton flux).  In Paper~I, the analysis was performed for the simple Leaky Box Model
(LBM) to validate the approach. We extend the analysis for the more realistic diffusion model,
by considering constraints set by radioactive nuclei. The model is the minimal reacceleration
one, with a constant Galactic wind perpendicular to the disc plane
\citep[e.g.,][]{2001ApJ...547..264J,2001ApJ...555..585M}, allowing for a central underdensity of gas (of a
few hundreds of pc) around the solar neighbourhood \citep{2002A&A...381..539D}.

The paper is organised as follows.
In Sect.~\ref{sec:PropagModel}, we recall the main ingredients of the diffusion model,
in particular the so-called local bubble feature. We briefly describe the MCMC
technique in Sect.~\ref{sec:MCMC} (the full description was given in Paper~I).
We then estimate the transport parameters in the 1D and 2D geometry. In
Sect.~\ref{sec:1D_BC}, this is performed at fixed $L$ (halo size of the Galaxy),
using the B/C ratio only. The analysis is extended in Sect.~\ref{sec:1D_rad} by taking
advantage of the radioactive nuclei to break the well-known degeneracy between the
parameters $K_0$ (normalisation of the diffusion coefficient) and $L$. 
We then present our conclusions in Sect.~\ref{sec:conclusion}. 

%%%%%%%%%%%%%%%%%%%%%%%%%%%%%%%%%%%%%%%%%%%%%%%%%%%%%%%%%%%%%%%%%%%%%%
%%%%%%%%%%%%%%%%%%%%%%%%%%%%%%%%%%%%%%%%%%%%%%%%%%%%%%%%%%%%%%%%%%%%%%
\section{Propagation model \label{sec:PropagModel}}

The set of $j=1\dots n$ equations governing the propagation of $n$ CR nuclei in the Galaxy is
described in \citet{1990acr..book.....B}. It is a generic diffusion/convection equation with
energy gains and losses. 
Depending on the assumptions made about the spatial and energy dependence of the transport
coefficients, semi-analytical or fully numerical procedures are necessary to solve this set of
equations. The solution also depends on the boundary conditions, hence on the geometry of the
model for the Galaxy. 

Several diffusion models are considered in the literature
\citep{1992ApJ...390...96W,1993A&A...267..372B,1998ApJ...509..212S,2001ApJ...547..264J,
2001ApJ...555..585M,2003A&A...410..189B,2006ApJ...642..882S,2008JCAP...10..018E,2008ApJ...681.1334F}.
We use a popular two-zone diffusion model with minimal reacceleration, where the Galactic
wind is constant and perpendicular to the Galactic plane.
The 1D and 2D version of this model are discussed, e.g., in \citet{2001ApJ...547..264J} and
\citet{2001ApJ...555..585M}. For the sake of legibility, the solutions are given in
Appendix~\ref{App:solutions}. 

Below, we reiterate the assumptions of the model, and describe the free parameters
that we constrain in this study (Sect.~\ref{sec:free_param}). 
 
                     %---------------%
\subsection{Transport equation\label{s:transport}}
The differential density $N^j$ of the nucleus $j$ is a function of the total energy $E$ and
the position $\vec{r}$ in the Galaxy.  Assuming a steady state, the transport equation can be
written in a compact form as 
\begin{equation}
{\cal L}^j N^j + \frac{\partial}{\partial E}\left( b^j N^j - c^j \frac{\partial N^j}{\partial E} \right) = {\cal S}^j\;.
\label{eq:CR}
\end{equation}
The operator ${\cal L}$ (we omit the superscript $j$) describes the diffusion
$K(\vec{r},E)$ and the convection $\vec{V}(\vec{r})$ in the Galaxy, but also the decay rate
$\Gamma_{\rm rad}(E)= 1/(\gamma\tau_0)$ if the nucleus is radioactive, and the destruction
rate $\Gamma_{\rm inel}(\vec{r},E)=\sum_{ISM} n_{\rm ISM}(\vec{r}) v \sigma_{\rm inel}(E)$ for
collisions with the interstellar matter (ISM), in the form
\begin{equation}
{\cal L}(\vec{r},E) =  -\vec{\nabla} \cdot (K\vec{\nabla}) + \vec{\nabla}\cdot\vec{V} +
     \Gamma_{\rm rad} + \Gamma_{\rm inel}.
\label{eq:operator}
\end{equation}

The coefficients $b$ and $c$ in Eq.~(\ref{eq:CR}) are respectively first and
second order gains/losses in energy, with
\begin{eqnarray}
\label{eq:b}
b\,(\vec{r},E)&=& \big\langle\frac{dE}{dt}\big\rangle_{\rm ion,\,coul.} 
   - \frac{\vec{\nabla}.\vec{V}}{3} E_k\left(\frac{2m+E_k}{m+E_k}\right)
	 \\\nonumber
   &  & + \;\; \frac{(1+\beta^2)}{E} \times K_{pp},\\
\label{eq:c}
c\,(\vec{r},E)&=&  \beta^2 \times K_{pp}.
\end{eqnarray}
In Eq.~(\ref{eq:b}), the ionisation and Coulomb energy losses are taken from
\citet{1994A&A...286..983M} and \citet{1998ApJ...509..212S}. The divergence of the Galactic
wind $\vec{V}$ gives rise to an energy loss term related to the adiabatic expansion of
cosmic rays. The last term is a first order contribution in energy from reacceleration.
Equation~(\ref{eq:c}) corresponds to a diffusion in momentum space, leading to an energy gain.
The associated diffusion coefficient
$K_{pp}$ (in momentum space) is taken from the model of minimal reacceleration by the
interstellar turbulence \citep{1988SvAL...14..132O,1994ApJ...431..705S}. It is related to the
spatial diffusion coefficient $K$ by
\begin{equation}
K_{pp}\times K= \frac{4}{3}\;V_a^2\;\frac{p^2}{\delta\,(4-\delta^2)\,(4-\delta)},
\label{eq:Va}
\end{equation}
where $V_a$ is the Alfv\'enic speed in the medium.

The source term ${\cal S}^j$ is a combination of i) primary sources
$q^j(\vec{r},E)$ of CRs (e.g., supernovae), ii) secondary fragmentation-induced sources 
$\sum_k^{m_k>m_j} n_{\rm ISM}(\vec{r}) v \sigma^{k\rightarrow j}_{\rm frag}(E)N^k(\vec{r},E)$,
and iii) secondary decay-induced sources $\sum_k N^k(\vec{r},E)/(\gamma\tau_0^{k\rightarrow
j})$. In particular, the secondary contributions link one species to all heavier nuclei,
coupling together the $n$ equations. However, the matrix is triangular and one possible
approach is to solve the equation starting from the heavier nucleus (which is always
assumed to be a primary).

                    %---------------%
\subsection{Geometry of the Galaxy and simplifying assumptions}
\begin{figure}[t]
\centering
\includegraphics[width = \columnwidth]{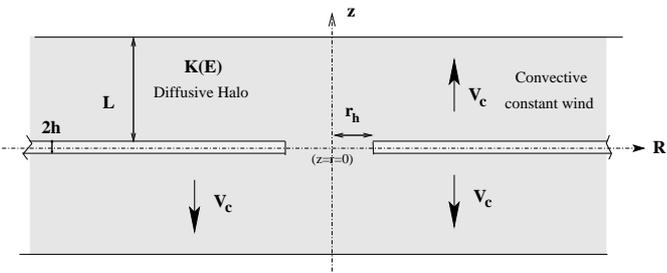}
\caption{Sketch of the model: sources and interactions (including energy losses and gains)
are restricted to the thin disc $\propto 2h\delta(r)$. Diffusion $K$ and convection $V_c$
transport nuclei in both the disc (half-height $h$) and the halo (half-height $L$). The Galaxy radial
extension is $R$. The local bubble is featured to be a cavity of radius $r_h$ in the disc 
devoid of gas.}
\label{fig:model}
\end{figure}

The Galaxy is modelled to be a thin disc of half-thickness $h$, which contains the gas and the
sources of CRs. This disc is embedded in a cylindrical diffusive halo of half-thickness $L$, where the gas
density is assumed to be 0. CRs diffuse into both the disc and the halo independently of their
position. A constant wind $V_c$ perpendicular to the Galactic plane is also considered.
This is summarised in Fig.~\ref{fig:model} (see next section for the definition of $r_h$). 

We use the $\delta(z)$ approximation introduced in \citet{1979ApJ...229..747J},
\citet{1990A&A...237..445P}, and \citet{1992ApJ...390...96W}. Considering the radial
extension $R$ of the Galaxy to be either infinite or finite leads to the 1D version or 2D version of
the model, respectively. The corresponding sets of equations (and their solutions) obtained after these
simplifications are presented in Appendix~\ref{App:solutions}.
These assumptions allow for semi-analytical solutions of the problem, as the interactions
(destruction, spallations, energy gain and losses) are restricted to the thin disc. The gain
is in the computing time, which is a prerequisite for the use of the MCMC technique, where
several tens of thousands of models are calculated. These semi-analytical models reproduce all
salient features of full numerical approaches (e.g., \citealt{1998ApJ...509..212S}), and
they are useful for systematically studying the dependence on key parameters, or some
systematics of the parameter determination \citep{2010A&A...xxx..xxxM}.

We note that most of the results of the paper are based on the 1D geometry (solutions only depend on
$z$), which is less time-consuming than the 2D one in terms of computing time\footnote{The 2D solution is
based on a Bessel expansion/resummation [see Eq.~(\ref{eq:Niresum})]. For each Bessel order,
an equation similar to that for the 1D geometry needs to be solved. Nine Bessel orders are in
many cases enough to ensure convergence \citep{2001ApJ...555..585M}, but at least 100 orders
are required  in the general case, which multiply the computing time by roughly the
same amount.}. The parameter degeneracy is also more easily extracted and
understood in this case~\citep{2001ApJ...547..264J,2006astro.ph.12714M}. Nevertheless,
the results for the 2D geometry are also reported, as it has been used in a series of
studies inspecting stable nuclei
\citep{2001ApJ...555..585M,2002A&A...394.1039M}, $\beta$-radioactive nuclei
\citep{2002A&A...381..539D}, standard anti-nuclei
\citep{2001ApJ...563..172D,2008PhRvD..78d3506D,2009PhRvL.102g1301D} and positrons
\citep{2008arXiv0809.5268D}. It has also been used to set constraints on dark matter
annihilations in anti-nuclei \citep{2004PhRvD..69f3501D}, and positrons
\citep{2008PhRvD..77f3527D}. The reader is referred to these papers, and especially
\citet{2001ApJ...555..585M} for more details and references about the 2D case.

                    %---------------%
\subsection{Radioactive species and the local bubble\label{sec:rad}}
Our model does not take into account all the observed irregularities of the gas distribution,
such as holes, chimneys, shell-like structures, and disc flaring. The main reason is that
as far as stable nuclei are concerned, only the average grammage crossed is relevant when
predicting their flux (which motivates LBM). As such, the thin-disc approximation is
a good trade-off between having a realistic description of the structure of the Galaxy and
simplicity.

However, the local distribution of gas affects the flux calculation of radioactive species
\citep{1997AdSpR..19..787P,1998A&A...337..859P,2002A&A...381..539D}.
We consider a radioactive nucleus that diffuses in an unbound volume and decays with a rate
$1/(\gamma\tau_0)$. In spherical coordinates, appropriate to describe this
situation, the diffusion equation reads
\begin{equation}
-K \triangle_{\vec{r}} G + \frac{G}{\gamma\tau_0} = \delta(\vec{r}) \;.
\end{equation}
The solution for the propagator $G$ (the flux is measured at $\vec{r}=\vec{0}$ for simplicity) is
\begin{equation}
G(\vec{r'})\propto \frac{e^{-r'/\sqrt{K\gamma\tau_0}}}{r'}\;.
\end{equation}
Secondary radioactive species, such as $^{10}$Be, originate from the spallations of the CR
protons (and He) with the ISM.  We model the source term to be a thin gaseous disc, except in a
circular region of radius $r_h$ at the origin. In the $\delta(z)$ approximation (see
Fig.~\ref{fig:model}) and in cylindrical coordinates,
\begin{equation}
Q(r,z)\propto \Theta(r-r_h) \delta(z) \;,
\end{equation}
where $\Theta$ is the Heaviside function.
The flux of a radioactive species is thus given by (we rewrite the propagator in
cylindrical coordinates)
\begin{equation}
  N(r\!=\!z\!=\!0) \propto \!\int_{0}^\infty \!\!\int_{-\infty}^{+\infty} \!\!\!\!\!G(\sqrt{r'^2\!+\!z'^2})\,Q(r',z')\, r'dr' dz'.
\end{equation}
The ratio of the flux calculated for a cavity/hole $r_h$ to that of the
flux without hole ($r_h=0$) is
\begin{equation}
\frac{N_{r_h}}{N_{r_h= 0}}=\exp\left(\frac{-r_h}{\sqrt{K\gamma\tau_0}}\right)
   =\exp\left(\frac{-r_h}{l_{\rm rad}}\right) \;.
	 \label{eq:rad_damping}
\end{equation}
The quantity $l_{\rm rad}=\sqrt{K \gamma \tau_0}$ is the typical distance on which a
radioactive nucleus diffuses before decaying. Using $K\approx 10^{28}$~cm$^2$~s$^{-1}$  and
$\tau\approx 1$~Myr, the diffusion length is $l_{\rm rad}\approx 200$ pc.
Hence, in principal, any underdensity on a scale $r_h\sim 100$~pc about the Sun leads to an
exponential attenuation of the flux of radioactive nuclei. This attenuation is both
energy-dependent and species-dependent. It is energy-dependent because it decreases with the energy
as the time-of-flight of a radioactive nucleus is boosted by both its Lorentz factor and
the increase in the diffusion coefficient. It is species-dependent because nuclei half-lives for
the standard $Z<30$ cosmic-ray clocks range from 0.307 Myr for $^{36}$Cl to 1.51 Myr for $^{10}$Be.

In this paper, we model the local bubble to be this simple hole in the gaseous disc, as shown
in Fig.~\ref{fig:model}. The exponential decrease in the flux of this modified DM,
as given by Eq.~(\ref{eq:rad_damping}), is directly plugged into the solutions for the 
standard DM ($r_h=0$). In principle, i) the
hole has also an impact on stable species as it decreases the amount of matter available for
spallations, and ii) in the 2D geometry, a hole at $R_\odot=8$~kpc breaks down the cylindrical geometry.
However, in practice, \citet{2002A&A...381..539D} found that the first effect is minor, and that
the hole can always be taken to be the origin of the Galaxy (the impact of the $R$ boundary being
negligible for radioactive species). 

Other subleties exist, which were not considered in \citet{2002A&A...381..539D}.
Indeed, the damping in the solar neighbourhood|combined with the production of the
radioactive species matching the data at low energy|means that at intermediate GeV/n energies,
the flux of this radioactive species is higher in the modified model (with $r_h\neq0$) than in the
standard one (with $r_h=0$). It also means that everywhere else in the Galactic disc, at all energies,
the radioactive fluxes are higher in the modified model (with damping). There are two consequences:
i) all spallative products from these radioactive nuclei originate in an effective 
diffusion region in the disc \citep{2003A&A...402..971T}, the size of which may be much larger than
the size of the underdense bubble. In this case, these products ought to be calculated from the undamped fluxes;
ii) the decay products of these radioactive nuclei (e.g., $^{10}$B, which originates
from the $\beta$-decay of $^{10}$Be) are stable species that originate in an effective
diffusion sphere (decay can occur not only in the disc, but in the halo).
Both these effects must be considered because their contributions potentially affect
the calculation, e.g., of the B/C and Be/B ratios (by means of the B flux), which are
used to fit the models.
We confirm that taking spallative products from the damped or undamped radioactive fluxes
left these ratios unchanged. On the other hand, for the decay products, the effect is of the
order of $1-10\%$, which is in general enough to change the values of the best-fit
parameters.
However, the {\em average} flux (over the effective diffusion zone) from which
the decay products originate lies between the damped and undamped values: the lower
the effective diffusive sphere, the closer the flux is to the damped one.
In particular, at low energy, when convection is allowed,
the diffusion zone can be small \citep{2003A&A...402..971T}. 

To keep the approach simple, we use here the damped flux
of radioactive species for all spallative and decay products (as was implicitly
assumed in \citealt{2002A&A...381..539D}). This approach is expected to provide
the maximal possible size for $r_h$ (if a non-null value is preferred by the fit).

                    %---------------%
 \subsection{Input ingredients and free parameters of the study\label{sec:free_param}}

                             %%%%%
\subsubsection{Gas density} The gas density scale height strongly varies with $r$
depending on the form considered|neutral, molecular, or ionised
(see, e.g., \citealt{2001RvMP...73.1031F}). We use the surface density measured in the
solar neighbourhood as a good estimate of the average gas in the Galactic disc. 
We set $n_{\rm ISM}=1$~cm$^{-3}$, which corresponds to a surface density $\Sigma_{\rm ISM}=2hn_{\rm
ISM}\sim6\times 10^{20}$~cm$^{-2}$ \citep{2001RvMP...73.1031F}. The number fraction of H and He is
taken to be 90\% and 10\%, respectively. The ionised-hydrogen space-averaged density may be identified with the
free-electron space-averaged density, which is the sum of the contributions of H$_{\rm II}$ regions
and the diffuse component \citep{2001AJ....122..908G,2001RvMP...73.1031F}. The intensity of the latter
is well measured  $0.018\pm0.002$~cm$^{-3}$ \citep{2006AN....327...82B,2008A&A...490..179B}, whereas
the former depends strongly on the Galactocentric radius $r$ \citep{2009ApJ...690..706A}.  For the
total electron density, we choose to set $\langle n_{e^-}\rangle=0.033$ and $T_e\sim10^4\unit{K}$
\citep{1992AJ....104.1465N}.

The disc half-height is set to be $h=100$~pc. It is not a
physical parameter {\em per se} in the $\delta(z)$ approximation, although it is related to the
phenomena occurring in the thin disc. Physical parameters are related to
the surface density, which is easily rescaled from that calculated setting
$h=100$~pc (should we use a different $h$ value). In the 2D geometry, the boundary
is set to be $R=20$~kpc and the sun is located at $R_\odot=8.0$~kpc.

                             %%%%%
\subsubsection{Fragmentation cross-sections} In Paper~I, the sets of fragmentation cross-sections were
taken from the semi-empirical formulation of \citet{1990PhRvC..41..566W} updated in
\citet{1998ApJ...508..940W} (see also \citealt{2001ApJ...555..585M} and references therein). In this
paper, they are replaced by the 2003 version, as given in \citet{2003ApJS..144..153W}.
Spallations on He are calculated with the parameterisation of \citet{1988PhRvC..37.1490F}.

                             %%%%%
\subsubsection{Source spectrum} We assume that a universal source spectrum for
all nuclei exists, and that it has a simple power-law description. As in Paper~I, we assume
that $Q(E)\propto
\beta^{\eta} R^{- \alpha}$. The parameter $\alpha$ is the spectral index of the sources
and $\eta$ encodes the behaviour of the spectrum at low energy.
The normalisations of the spectra are given by the source abundances $q_j$, which are renormalised
during the propagation step to match the data at a specified kinetic energy per nucleon (usually
$\sim 10$~GeV/n). The correlations between the source and the transport parameters and their
impact on the transport parameter determination were discussed in Paper~I. In this study,
we set $\eta=-1$ and $\gamma=\alpha+\delta=2.65$ \citep{2008ApJ...678..262A}. Constraints
on the source spectra from the study of the measured primary fluxes are left to a subsequent
paper (Donato et al., in preparation).

                             %%%%%
\subsubsection{Free parameters} 

We have two geometrical free parameters
\begin{itemize}
   \item $L$, the halo size of the Galaxy (kpc);
   \item $r_h$, the size of the local bubble (kpc), which is most of the time set to be 0 (to compare with
	 models in the literature that do not consider any local underdensity);
\end{itemize}
and four transport ones
\begin{itemize}
   \item $K_0$, the normalisation of the diffusion coefficient (in unit of kpc$^2$~Myr$^{-1}$);
   \item $\delta$, the slope of the diffusion coefficient;
   \item $V_c$, the constant convective wind perpendicular to the disc (km~s$^{-1}$);
   \item $V_a$, the Alfv\'enic speed (km~s$^{-1}$) regulating the reacceleration strength [see Eq.~(\ref{eq:Va})].
\end{itemize}
The diffusion coefficient is taken to be 
\begin{equation}
K(E)= \beta K_0 {\cal R}^{\delta}.
\label{eq:K(E)}
\end{equation}

%%%%%%%%%%%%%%%%%%%%%%%%%%%%%%%%%%%%%%%%%%%%%%%%%%%%%%%%%%%%%%%%%%%%%%
%%%%%%%%%%%%%%%%%%%%%%%%%%%%%%%%%%%%%%%%%%%%%%%%%%%%%%%%%%%%%%%%%%%%%%
\section{MCMC \label{sec:MCMC}}

The MCMC method, based on the Bayesian statistics, is used here to estimate the 
full distribution|the so-called conditional probability-density function (PDF)|given
some experimental data (and some prior density for these parameters). We summarise
below the salient features of the MCMC technique. A detailed description of the
method can be found in Paper~I. The issue of the efficiency, which was not raised
in Paper I, is discussed in Appendix~\ref{sec:MCMC_combi}.

The Bayesian approach aims to assess the extent to which an experimental dataset
improves  our knowledge of a given theoretical model. Considering a model depending on
$m$ parameters 
\begin{equation}
\vectheta\equiv \{ \theta^{(1)},\,\theta^{(2)}, \, \ldots, \,\theta^{(m)}\},
\label{eq:theta}
\end{equation}
we wish to determine the PDF of the parameters given the data, $P (\vectheta|\mathrm{data})$. This so-called 
{\em posterior} probability quantifies the change in the degree of belief one 
can have in the $m$ parameters of the model in the light of the data.
Applied to the parameter inference, Bayes theorem reads
\begin{equation}
P (\vectheta|\mathrm{data}) =  \frac{P (\mathrm{data}|\vectheta)
\cdot P (\vectheta)}{P (\mathrm{data})}, 
\label{eq:bayes_theorem}
\end{equation}
where $P(\mathrm{data})$ is the data probability (the latter does not depend on the
parameters and hence, can be treated as a normalisation factor). 
This theorem links the posterior probability to the
likelihood of the data ${\cal L(\vectheta})\equiv P (\mathrm{data}|\vectheta)$
and the so-called {\em prior} probability, $P (\vectheta)$, which indicates the degree 
of belief one has {\em before} observing the data.
The technically difficult point of Bayesian parameter estimates lies in the
determination of the individual posterior PDF, which requires an (high-dimensional) 
integration of the overall posterior density. Thus an efficient sampling
method for the posterior PDF is mandatory.

In general, MCMC methods attempt to studying any {\em target} distribution of a vector of
parameters, here  $P (\vectheta|\mathrm{data})$, by generating a sequence of
$n$ points/steps (hereafter a chain)
\begin{equation}
\{\vec{\theta}_i\}_{i=1, \ldots, n}\equiv \{ \vectheta_1,\,\vectheta_2, \, \ldots, \,\vectheta_n\}.
\end{equation} 
Each $\vectheta_i$ is a vector of $m$ components, e.g., as defined in
Eq.~(\ref{eq:theta}). In addition, the chain is Markovian in the sense that
 the distribution of $\vectheta_{n+1}$ is influenced entirely
by the value of $\vectheta_n$. MCMC algorithms are developed to ensure that the time
spent by the Markov chain in a region of the parameter space is proportional to 
the target PDF value in this region. 
Here, the prescription used to generate the Markov chains is the so-called Metropolis-Hastings algorithm,
which ensures that the stationary distribution of the chain asymptotically
tends to the target PDF. 

The chain analysis is based on the selection of a subset of points from the chains
(to obtain a reliable estimate of the PDF). Some steps at the beginning of the chain
are discarded (burn-in length). By construction, each step of the chain is correlated
with the previous steps: sets of independent samples are obtained by thinning the
chain (over the correlation length). The fraction of independent samples measuring
the efficiency of the MCMC is defined to be the fraction of steps remaining after
discarding the burn-in steps and thinning the chain.
The final results of the MCMC analysis are the target PDF and all marginalised PDFs.
They are obtained by merely counting the number of samples within the related
region of parameter space.

%%%%%%%%%%%%%%%%%%%%%%%%%%%%%%%%%%%%%%%%%%%%%%%%%%%%%%%%%%%%%%%%%%%%%%
%%%%%%%%%%%%%%%%%%%%%%%%%%%%%%%%%%%%%%%%%%%%%%%%%%%%%%%%%%%%%%%%%%%%%%
\section{Results for stable species (fixed halo size $L$)\label{sec:1D_BC}}

For stable species, the degeneracy between the normalisation of the diffusion coefficient
$K_0$ and the halo size of the Galaxy $L$ prevents us from being able to constrain both parameters
at the same time. We choose to set $L=4$~kpc (we also set $r_h=0$,
i.e., standard DM). The free transport parameters are $\{K_0,\,\delta,\,V_c,\,V_a\}$.
The classes of models considered are summarised in Table \ref{tab:class_models}.
\begin{table}[!t]
\caption{Classes of models tested in the paper.}
\label{tab:class_models}
\centering
\begin{tabular}{ccc} \hline\hline
Model    & $\!\!\!$Transport parameters$\!\!\!$  & Description  \\
\hline\vspace{-0.2cm}
& \multicolumn{2}{c}{} \\ 
I        & $\{K_0,\, \delta, \, V_c\}$         & Diffusion + convection   \vspace{0.05cm}\\
II       & $\{K_0,\, \delta, \, V_a\}$         & Diffusion + reacceleration \vspace{0.05cm}\\
III      & $\{K_0,\, \delta, \, V_c, \, V_a\}$ & Diff. + conv. + reac.    \vspace{0.cm}\\
\hline
\end{tabular}
\end{table}
The reference B/C dataset (denoted dataset F) used for the analysis is
described in Appendix~\ref{datasetBC}.

            %%#######################################%%
     \subsection{PDF for the transport parameters}

\begin{table}[!tb]
\caption{Most probable values for B/C data only ($L=4$~kpc).}
\label{tab:DM_results}\centering
\begin{tabular}{lcccc} \hline\hline
$\!$Model$\!$ & $K_{0} \times 10^2$   &  $\delta$  &    $V_{c}$  &  $V_a$ \\ 
Data    & $\!\!\!\!\!\!\!\!\!\!(\unit[]{kpc^2Myr^{-1}})\!\!\!\!\!\!\!\!\!\!$  &    &  $\!\!\!\!\!\!\!\!\!(\unit[]{km\,s^{-1})}\!\!\!\!\!\!\!\!\!$ & $\!\!\!\!\!\!\!\!\!\!\!(\unit[]{km\,s^{-1}})\!\!\!\!\!\!\!\!\!\!\!$ \\\hline
& \multicolumn{4}{c}{} \\ 
I-F     & $0.42^{+0.03}_{-0.04}$ & $0.93^{+0.02}_{-0.03}$ & $13.5^{+0.3}_{-0.3}$ & $\cdots$  \vspace{0.1cm}\\
II-F    & $9.7^{+0.3}_{-0.2}$  & $\!\!\!\!\!0.234^{+0.006}_{-0.005}\!\!\!\!\!$ & $\cdots$ & $73^{+2}_{-2}$ \vspace{0.1cm}\\
III-F   & $0.46^{+0.08}_{-0.06}$  & $0.86^{+0.04}_{-0.04}$ &  $18.9^{+0.3}_{-0.4}$ & $38^{+2}_{-2}$ \vspace{0.1cm}\\
\hline
\end{tabular}
\end{table}

\begin{figure}[!ht]
\includegraphics[width = 0.715\columnwidth]{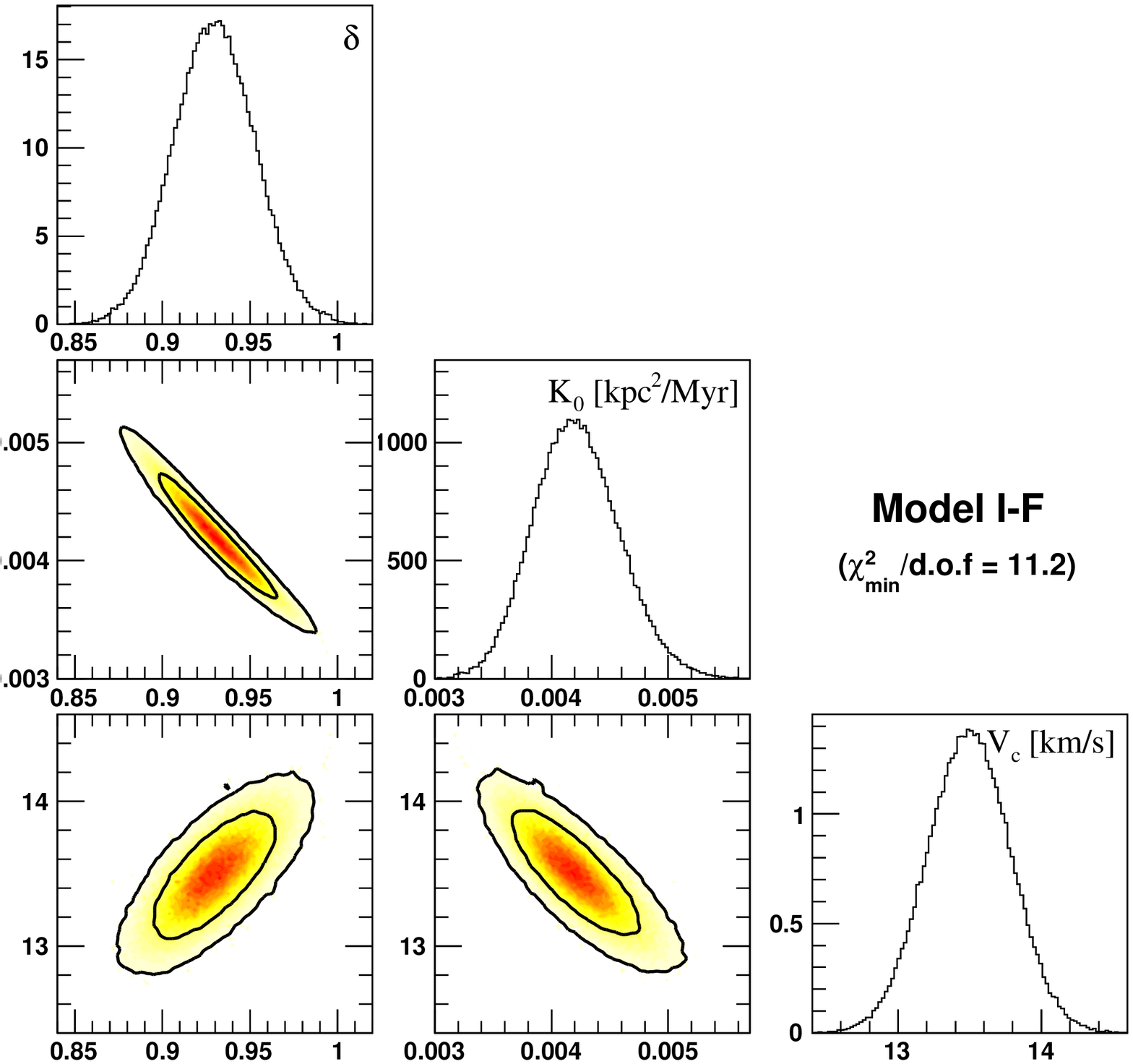}
\includegraphics[width = 0.715\columnwidth]{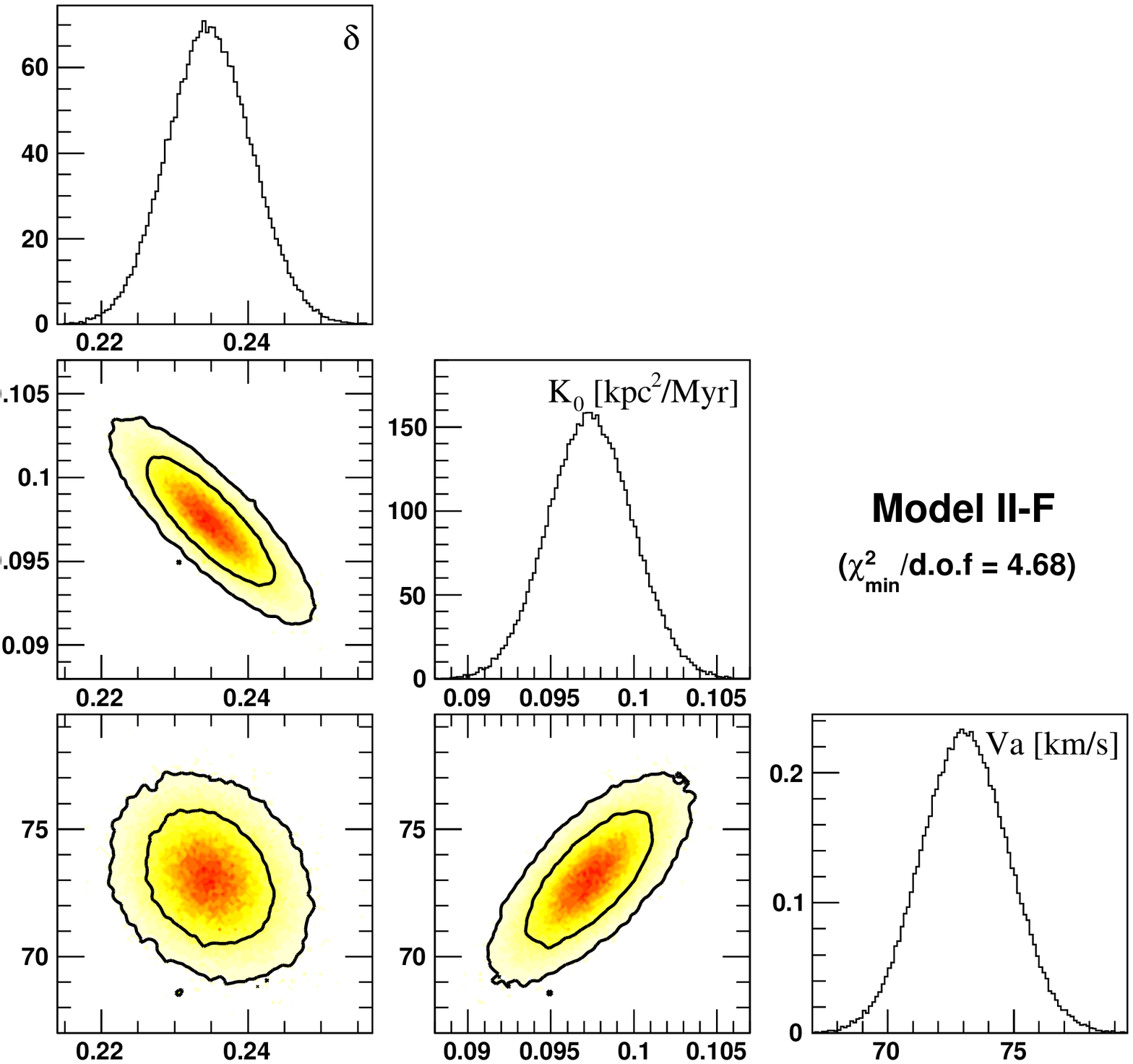}
\includegraphics[width = 0.95\columnwidth]{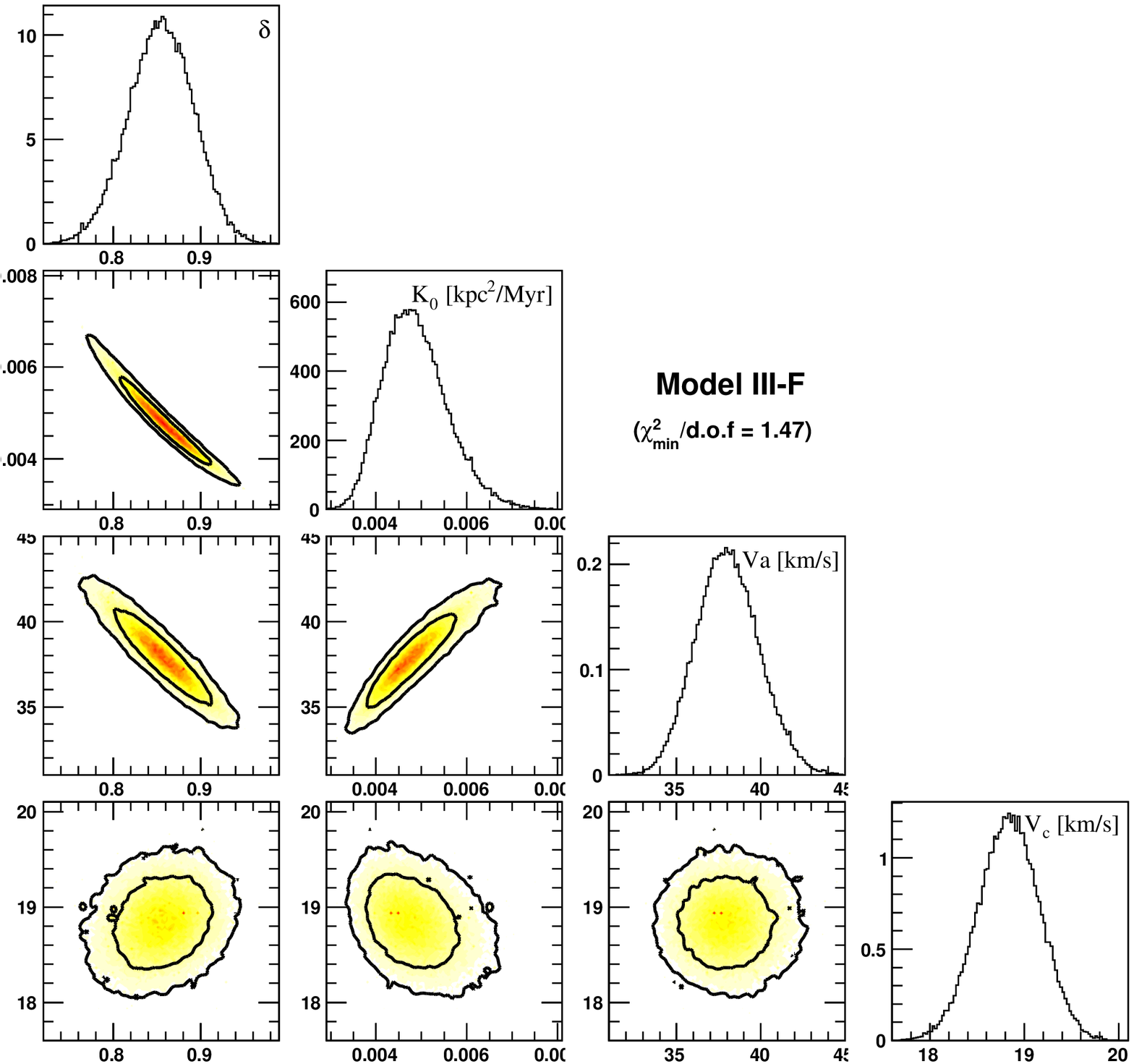}
\caption{From top to bottom: posterior PDFs of models I, II, and III using the B/C constraint (dataset F).
The diagonals show the 1D marginalised PDFs of the indicated parameters. Off-diagonal plots
show the 2D marginalised posterior PDFs for the parameters in the same column and same line respectively.
The colour code corresponds to the regions of increasing probability (from paler to darker shade), and the
two contours (smoothed) delimit regions containing, respectively, 68\% and 95\% (inner and outer contour) of
the PDF.}
\label{fig:MCMC_I+II+III}
\end{figure}

We begin with the PDFs of the parameters based on the B/C constraint (dataset F)
for the various classes of models (I, II, or III). The PDFs are shown in Fig.~\ref{fig:MCMC_I+II+III}. 

The first important feature is that the marginal distributions of the transport
parameters (diagonals) are mostly Gaussian. From the off-diagonal distributions, we
remark that $K_0$ and $\delta$ are negatively correlated. This originates in the
low-energy relation $K(E)\propto K_0 R^\delta$, which should remain approximately
constant to reproduce the bulk of the data at GeV/n energy. The diffusion slope
$\delta$ is negatively correlated with $V_a$, which is related to a
smaller $\delta$ being obtained if more reacceleration is included. On the other hand,
the positive correlation between $\delta$ and $V_c$ indicates that larger
$\delta$ are expected for larger wind velocities.

We show in Table~\ref{tab:DM_results} the most probable values of the transport
parameters, as well as their uncertainties, corresponding to 68\% confidence levels (CL)
of the marginalised PDFs. The precision to which the parameters are obtained is excellent,
ranging from a few \% to 10\% at most (for the slope of the diffusion coefficient $\delta$ in III).
This corresponds to statistical uncertainties only. These uncertainties are
of the order of, or smaller than systematics generated from uncertainties in the
input ingredients \citep[see details in][]{2010A&A...xxx..xxxM}.

As found in previous studies \citep[e.g.,][]{2005JCAP...09..010L}, for pure
diffusion/reacceleration models (II), the value of the diffusion slope $\delta$ found
is low ($\approx 0.23$ here). When convection is included (I and III), $\delta$ is large
($\approx 0.8-0.9$). This scatter in $\delta$ was already observed in \citet{2001ApJ...547..264J},
who also studied different classes of models. The origin of this scatter is consistent with the
aforementioned correlations in the parameters \citep[see also][]{2010A&A...xxx..xxxM}.

\begin{figure}[t]
	\centering
	\includegraphics[width = 0.48\textwidth]{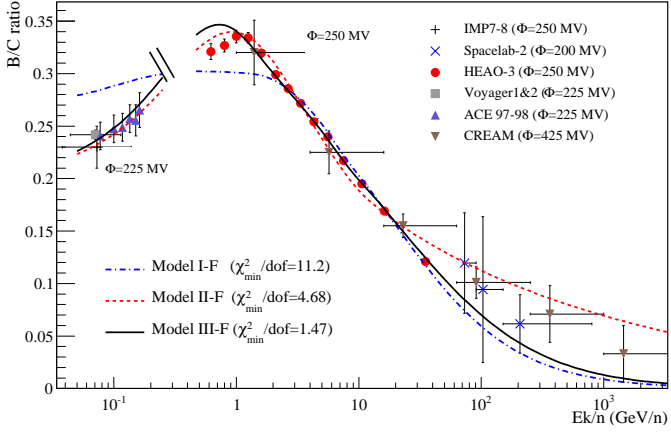}
  \caption{Best-fit ratio for model I (blue-dotted line), II (red-dashed line), and
model III (black-solid line) using dataset~F: IMP7-8, Voyager1\&2, ACE-CRIS, HEAO-3, Spacelab, and CREAM. The curves are
modulated with $\Phi = \unit[250]{GV}$ (and $\Phi = \unit[225]{GV}$ at low
energy). The corresponding best-fit parameters are gathered
in Table~\ref{tab:best_fitBC}.}
\label{fig:BC}
\end{figure}

The best-fit model parameters (which are not always the most probable ones) are given in
Table~\ref{tab:best_fitBC}, along with the minimal $\chi^2$ value per degree of freedom,
$\chi_{\mathrm{min}}^2/$d.o.f (last column). As found in previous analyses 
\citep{2001ApJ...555..585M,2002A&A...394.1039M}, the DM with both reacceleration
and convection reproduces the B/C data more accurately than without:
$\chi^2$/d.o.f$=1.47$ for III, 4.90 for II, and 11.6 for I.
\begin{table}[t]
\caption{Best-fit model parameters for B/C data only ($L=4$~kpc).}
\label{tab:best_fitBC}
\centering
\begin{tabular}{lcccccc} \hline\hline
Model    & $\!\!K_0^{\rm best}\times 10^2\!\!$     & $\delta^{\rm best}$ &  $V_c^{\rm best}$  &  $V_a^{\rm best}$  & $\!\!\!\!\!\!\chi^2$/d.o.f$\!\!\!$   \\
Data        & $\!\!\!\!\!\!$(kpc$^2\,$Myr$^{-1}$)$\!\!\!\!\!\!$ &        &  (km$\;$s$^{-1}$)  & (km$\;$s$^{-1}$)   & \\\hline
& \multicolumn{5}{c}{} \\ 
I-F  $\!\!\!\!\!\!$& 0.42 & 0.93 & 13.5 & \dots&  11.2  \vspace{0.05cm}\\  %chi2=313.467 (31)
II-F $\!\!\!\!\!\!$& 9.74 & 0.23 & \dots& 73.1 &  4.68  \vspace{0.05cm}\\  %chi2=132.499 (31)
III-F$\!\!\!\!\!\!$& 0.48 & 0.86 & 18.8 & 38.0 &  1.47  \vspace{0.15cm}\\  %chi2=39.7026 (31)
% N.B.: 161 steps between 0.05 and 5000 GeV/n used
\hline
\end{tabular}
\end{table}
The B/C ratio associated with these optimal $\chi^2$ values are displayed with the data
in Fig.~\ref{fig:BC}. We note that the {\em poor} fit for II (compared to III) is
explained by the departure of the model prediction from high-energy HEAO-3 data.

            %%#######################################%%
	\subsection{Sensitivity to the choice of the B/C dataset\label{sec:sensData}}

For comparison purposes, we now focus on several datasets for the B/C data. Low-energy
data points include ACE data, taken during the solar minimum period 1997-1998
\citep{2006AdSpR..38.1558D}. Close to submission of this paper, another ACE analysis was
published \citep{2009ApJ...698.1666G}. The 1997-1998 data points were reanalysed and
complemented with data taken during the solar maximum period 2001-2003. The AMS-01 also
provided B/C data covering almost the same range as the HEAO-3 data \citep{2010ICRC0182}.
Hence, for this section only, we attempt to analyse other B/C datasets that include these
components:
\begin{itemize}  
  \item A: HEAO-3 [$0.8-40$~GeV/n], 14 data points;
  \item C: HEAO-3 $\!\!+\!\!$ low energy [$0.3\!-\!0.5\!$~GeV/n], 22 data points;
  \item F: HEAO-3 $\!+\!$ low $\!+\!$ high energy [$0.2-2$~TeV/n], 31 data points;
  \item G1: as F, but with new ACE 1997-1998 data, 31 data points;
  \item G2: as F,  but with new ACE 2001-2003 data only, 31 data points;
  \item G1/2: using both 1997-1998 and 2001-2003 ACE data, 37 data points;
  \item H: as F, but HEAO-3 replaced by AMS-01 data, 27 data points.
\end{itemize}

\begin{figure}[t]
\centering
\includegraphics[width = \columnwidth]{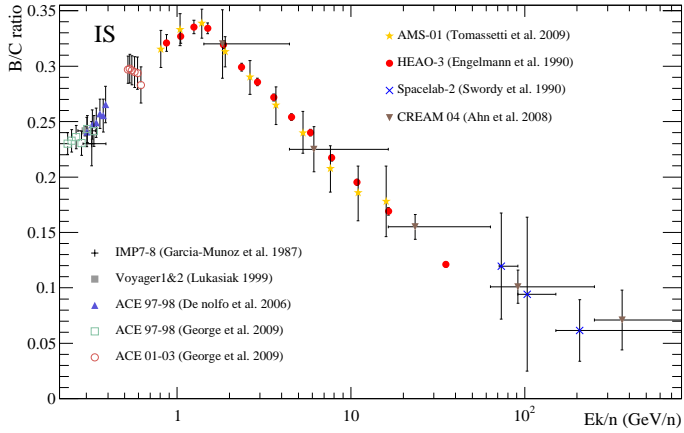}
\caption{B/C data used in this section. Shown are the IS data (rescaled from TOA data
using $E_k^{IS} = E_k^{\rm TOA} + \Phi$, see Paper~I). For several experiments,
in addition to the error bars in the ratio, we display the energy interval from which the
central energy point is obtained.}
\label{fig:BC_datasets}
\end{figure}

The data are shown in Fig.~\ref{fig:BC_datasets}. Thanks to the high level of modulation
for the 2001-2003 ACE data, the IS (demodulated) B/C ratio covers nicely the gap between
HEAO-3 and lower energy data. HEAO-3 and AMS-01 data also show consistency across
their whole energy range .

The best-fit model parameters for these data are shown in
Table~\ref{tab:different_data}. 
\begin{table}[t]
\caption{Best-fit model parameters based on different B/C datasets.}
\label{tab:different_data}
\centering
\begin{tabular}{lcccccc} \hline\hline
Model    & $\!\!K_0^{\rm best}\times 10^2\!\!$     & $\delta^{\rm best}$ &  $V_c^{\rm best}$ &  $V_a^{\rm best}$  & $\!\!\!\!\chi^2$/d.o.f$\!\!\!\!$   \\
Data        & $\!\!\!\!\!\!$(kpc$^2\,$Myr$^{-1}$)$\!\!\!\!\!\!$ &        &  (km$\;$s$^{-1}$)  & (km$\;$s$^{-1}$)   & \\\hline
& \multicolumn{5}{c}{} \\ 
III-A  $\!\!\!\!\!\!$& 2.51 & 1.00 & 21.7 & 35.4 &  2.11  \vspace{0.05cm}\\  %chi2=21.1207 (14)
III-C  $\!\!\!\!\!\!$& 0.43 & 0.89 & 18.9 & 36.7 &  1.72  \vspace{0.05cm}\\  %chi2=30.9401 (22)
III-F  $\!\!\!\!\!\!$& 0.48 & 0.86 & 18.8 & 38.0 &  1.47  \vspace{0.05cm}\\  %chi2=39.7026 (31)
III-G1 $\!\!\!\!\!\!$& 0.53 & 0.84 & 18.0 & 37.4 &  1.80  \vspace{0.05cm}\\  %chi2=48.5387 (31)
III-G2 $\!\!\!\!\!\!$& 0.46 & 0.85 & 20.0 & 39.6 &  2.73  \vspace{0.05cm}\\  %chi2=73.915  (31)
III-G1/2$\!\!\!\!\!\!$&0.53 & 0.83 & 19.0 & 39.1 &  2.94  \vspace{0.05cm}\\  %chi2=100.046 (37)
III-H  $\!\!\!\!\!\!$& 1.85 & 0.51 & 18.1 & 54.1 &  0.25  \vspace{0.15cm}\\  %chi2=5.69848 (27)
% N.B.: 161 steps between 0.05 and 5000 GeV/n used
\hline
\end{tabular}
\end{table}
The low-energy data play an important part in the fitting procedure: $\delta$ decreases by 0.1
when going from III-A to III-C, and the diffusion normalisation is decreased. When the CREAM
data at higher energy are taken into account (III-F), the best-fit diffusion slope $\delta$ again
becomes slightly lower (from 0.89 to 0.86), but CREAM data uncertainty is still too important to be
conclusive. The impact of the low-energy ACE reanalysed data points is seen when comparing
III-F with III-G1: the scatter between the derived best-fit parameters is already of the order
of the statistical uncertainty (see Table~\ref{tab:DM_results}). The data taken either during
the solar minimum period (G1) or the solar maximum period (G2) cover a different energy
range (see Fig.~\ref{fig:BC_datasets}). The $\chi^2_{\rm min}$ for G2 is greater, which
is not surprising, given the abnormal trend followed by these data (empty circles
in Fig.~\ref{fig:BC_datasets}). Nevertheless, it is reassuring to see that they lead
to consistent values of the transport parameters.

If we now replace the HEAO-3 data with the AMS-01 data, the impact
on the fit is striking: the best-fit diffusion slope $\delta$ goes from 0.86 to 0.51.
As discussed in \citet{2010A&A...xxx..xxxM}, HEAO-3 data strongly constrain the slope towards
$\delta \approx 0.8$, even if there is a systematic
energy bias in the HEAO-3 data themselves. From the AMS-01 data, we see that there could be a way
of reconciling the presence of a Galactic wind and reasonable values of $\delta$. However, the large error bars
in AMS-01 data, reflected by the low $\chi^2$/d.o.f value, does not allow to draw firm conclusions.
Data in the same energy range from PAMELA would be helpful in that respect. Moreover,
high energy data from subsequent CREAM flights or from the TRACER experiments
will be a crucial test of the diffusion slope: at TeV energies, diffusion
alone is expected to shape the observed spectra, so that the ambiguity with the effect of
convection or reacceleration is lifted \citep{2005APh....24..146C}.

  \subsection{Comparison of trends for the DM and for the LBM}

For completeness, we briefly comment on the similarities and differences
between the results found here and in Paper~I. To follow the organisation of the
previous sections, the comparison with the LBM is discussed for different
classes of models (I, II, and III), and then for different datasets (A, B, and C).
We note that the best-fit values presented below differ slightly for those given in Paper~I,
as an updated set of production cross-section is used.

We recall that in the LBM (see Paper~I), the free parameters are the normalisation of the
escape length $\lambda_0$, $\delta$, a cut-off rigidity $R_0$, and a pseudo-Alfv\'enic
speed ${\cal V}_a$. The latter is linked to a true speed by means of $V_a= {\cal V}_a \times
(hL)^{1/2}$, i.e., $V_a= 0.4^{1/2}\; {\cal V}_a$ for $h=0.1$~kpc and $L=4$~kpc. The
diffusion coefficient at 1 GV is related to the escape length by means of $K_0\approx 0.5\,
c \times \bar{\mu}L/\lambda_0$, where we use $\mu=2hn\bar{m}=1.34\times
10^{-3}$~g~cm$^{-2}$, leading to $K_0~($kpc$^2$~Myr$^{-1})\approx
0.82/\lambda_0~($g~cm$^{-2})$. The LBM parameters gathered in Table~\ref{tab:compare}
are obtained from the above conversions, to ease the comparison with the DM
results.
\begin{table}[t]
\caption{Best-fit parameters on B/C data for the LBM.}
\label{tab:compare}
\centering
\begin{tabular}{lcccccc} \hline\hline
Model    & $\!\!K_0^{\rm best}\times 10^2\!\!$     & $\delta^{\rm best}$ &  $R_0$  &  $V_a^{\rm best}$  & $\!\!\!\!\!\!\chi^2$/d.o.f$\!\!\!$   \\
Data        & $\!\!\!\!\!\!$(kpc$^2\,$Myr$^{-1}$)$\!\!\!\!\!\!$ &        &  (GV)   & (km$\;$s$^{-1}$)   & \\\hline
& \multicolumn{5}{c}{} \\ 
%
% I-A (Paper~I)   & 54.7 & 0.702& 4.21 & \dots & 3.35
% II-A (Paper~I)  & 25.8 & 0.51 &\dots & 88.8  & 1.43
% III-A (Paper~I) & 31.7 & 0.56 & 2.73 & 73.0  & 1.30
I-F  $\!\!\!\!\!\!$& 2.36 & 0.56 & 5.70 & \dots&  5.52  \vspace{0.05cm}\\  %chi2=149.151 (31)  lambda0=34.8
II-F $\!\!\!\!\!\!$& 5.26 & 0.38 & \dots& 65.1 &  1.78  \vspace{0.05cm}\\  %chi2=47.9904 (31)  lambda0=15.6  Pseudo-Va=103.
III-F$\!\!\!\!\!\!$& 4.19 & 0.43 & 2.94 & 53.9 &  1.56  \vspace{0.15cm}\\  %chi2=42.1351 (31)  lambda0=19.6  Pseudo-Va=85.2
% N.B.: 161 steps between 0.05 and 5000 GeV/n used
%%%%%
% III-C using WKS98 (Paper~I)    & 26.9 & 0.527& 2.45 & 88.5 & 1.06 \\ %chi2=18.0200 (22)
% III-C using WKS 98 (this paper)& 26.5 & 0.52 & 2.56 & 84.7 & 1.03 \\ %chi2=18.5513 (22)
% The slight different between the two results lies in a slight change of the normalisation routine to the data
%%%%%
III-A  $\!\!\!\!\!\!$& 2.32 & 0.57 & 4.40 & 11.3 &  2.71  \vspace{0.05cm}\\  %chi2=27.1086 (14)  lambda0=35.42 Pseudo-Va=17.92
III-C  $\!\!\!\!\!\!$& 4.13 & 0.44 & 3.10 & 53.6 &  2.26  \vspace{0.05cm}\\  %chi2=40.7154 (22)  lambda0=19.87 Pseudo-Va=84.8 
III-F  $\!\!\!\!\!\!$& 4.19 & 0.43 & 2.94 & 53.9 &  1.56  \vspace{0.05cm}\\  %chi2=42.1348 (31)  lambda0=19.6  Pseudo-Va=85.4
%III-G1 $\!\!\!\!\!\!$& 3.86 & 0.45 & 3.29 & 49.6 &  1.67  \vspace{0.05cm}\\  %chi2=45.1262 (31)  lambda0=21.26 Pseudo-Va=78.36
%III-G2 $\!\!\!\!\!\!$& 4.18 & 0.43 & 3.28 & 54.9 &  2.08  \vspace{0.05cm}\\  %chi2=56.0739 (31)  lambda0=19.61 Pseudo-Va=86.84
%III-G1/2$\!\!\!\!\!\!$&3.91 & 0.45 & 3.52 & 51.2 &  3.03  \vspace{0.15cm}\\  %chi2=64.4072 (37)  lambda0=20.97 Pseudo-Va=80.89
% N.B.: 161 steps between 0.05 and 5000 GeV/n used
\hline
\end{tabular}
\end{table}

For the different classes of models (I, II and III), a comparison of
Table~\ref{tab:best_fitBC} with the first three rows of Table~\ref{tab:compare} indicates
that the same trend is found. For instance, model~I (without reacceleration)
has a larger $\delta$ than those with, and model~II (without convection/rigidity-cutoff)
has a smaller $\delta$ than those with. The slope for model~III (with both convection
and reacceleration) is in-between. This effect is more marked for
the DM than for the LBM. We note that model~II (with reacceleration but without
convection) is almost consistent with a Kolmogorov spectrum of turbulence,
but is inconsistent with the data.
Concerning the different datasets (A, C, and F), again, the same trend as for the LBM is found
(compare Table~\ref{tab:different_data} and the last three rows of Table~\ref{tab:compare}).

The most striking difference between the two models (LBM and DM) concerns their $\delta$ values.
This difference can be explained in terms of non-equivalent parameterisation of
the low-energy transport coefficient (see \citealt{2010A&A...xxx..xxxM} for more
details). Apart from this, both the value of the Alfv\'enic speed and the normalisation
of the diffusion coefficient $K_0$ in the two cases are fairly consistent when
similar values of $\delta$ are considered.

  \subsection{Dependence of the parameters with $L$\label{sec:Lfree}}

\begin{figure}[t]
\centering
\includegraphics[width = \columnwidth]{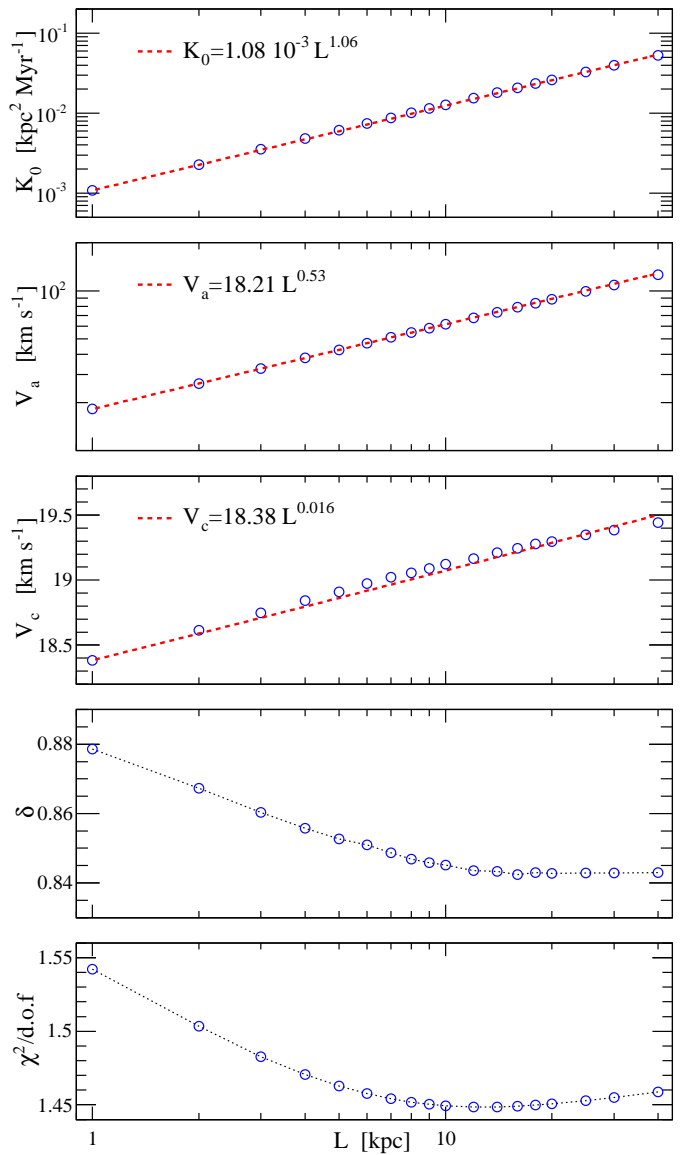}
\caption{Best-fit parameters (III-F) as a function of the halo size of the Galaxy (blue circles).
From top to bottom: $K_0$, $V_a$, $V_c$, $\delta$, and the associated $\chi^2_{\rm min}$.
In the first three figures, a parametric function matching the observed dependence
is shown (dashed-red line).}
\label{fig:BCparam_vs_L}
\end{figure}

All the previous conclusions were derived for $L=4$~kpc, but hold for any 
other halo size. The evolution of the transport parameters with $L$ is shown in Fig.~\ref{fig:BCparam_vs_L}
(the best-fit values are consistent with those found in \citealt{2002A&A...394.1039M}).
In the three upper figures, we have superimposed the observed dependence a parametric formula. 

For $K_0$ (top panel), the formula can be understood if we consider the grammage of the DM. 
In the purely diffusive regime, we have $\lambda_{\rm esc}\propto L/K$.
This means that when we vary $L$, to keep the same grammage in the {\em equivalent} LBM, we need
to vary $K_0$ accordingly. We find that $K_0=1.08\times 10^{-3} (L/1~{\rm kpc})^{1.06}$~kpc$^2$~Myr$^{-1}$
instead of $K_0\propto L$. The origin of the residual $L^{1.06}$ dependence is unclear. It may come
from the energy loss and gain terms. 

For the reacceleration, the interpretation is also simple. From Eq.~(\ref{eq:Va}), $V_a$ should
scale as $\sqrt{K_0}$, so that $V_a\propto \sqrt{L}$. We find $V_a=18.21 (L/1~{\rm kpc})^{0.53}$~km~s$^{-1}$.
This is exactly $V_a \propto\sqrt{K_0}$, with the dependence $\sqrt{L^{0.06}}$ as above.
The quantities $V_c$ and $\delta$ are roughly constant with $L$. The $\chi^2_{\rm min}$ surface
is rather flat, although a minimum is observed around $L\sim15$~kpc
(the presence of a minimum may be related to the presence of
the decayed $^{10}$Be into $^{10}$B in the B/C ratio). 
This flatness is a consequence of the degeneracy of $K_0/L$ when only stable
species are considered.
 Consequently, an MCMC with $L$ as an additional free parameter does not
converge to the stationary distribution. A sampling of the Galactic halo size is possible if
radioactive nuclei are considered to lift the above degeneracy (see
Sect.~\ref{sec:1D_rad}).

\subsection{Summary of stable species and generalisation to the 2D geometry\label{sec:summary_stable}}

The transport parameters for both LBM (Paper~I) and 1D DM, when fitted to
existing B/C data, are consistent with both convection and reacceleration. The correlations between
the various transport parameters, as calculated from the MCMC technique, are consistent with
what is expected from the relationships between DMs and the LBM
\citep[e.g.,][]{2006astro.ph.12714M}. From the B/C analysis point of view, 
it implies that even if we are unable to reach conclusions about the value
of $\delta$ (see \citealt{2010A&A...xxx..xxxM}), once
this value is known, all other transport parameters are well constrained.

The conclusions obtained for the 1D DM naturally hold for the 2D DM. We recall that the main
difference between the 1D and 2D geometry is that i) the spatial distribution of sources, which
was constant in 1D, is now $q(r)$; and ii) the Galaxy has a side-boundary at a radius taken to
be $R=20$~kpc. 
As a check, we first used the 2D solution (presented in Appendix~\ref{app:2D}) with
$R=20$~kpc, but set $q(r)$ to be constant. The best-fit parameters were in agreement with those
obtained from the 1D solution. We present in Table~\ref{tab:1Dvs2D} the best-fit parameters
for models II and III for $L=4$~kpc in the 2D solution where $q(r)$ follows the SN remnant
distribution of \citet{1998ApJ...504..761C}. The values for the 1D solution are also reported
for the sake of comparison. The main difference is in the value of $K_0$, which varies
by $\sim 10\%$ and also affects $V_a$ (by means of the ratio $V_a/\sqrt{K_0}$, which
is left unaffected). This is consistent with the variations found by \citet{2002A&A...394.1039M}.

%\begin{table}[!tb]
%\caption{Most probable values for B/C data only ($L=4$~kpc) for the 2D DM.}
%\label{tab:DM2D_results}\centering
%\begin{tabular}{lcccc} \hline\hline
%$\!$Model$\!$ & $K_{0} \times 10^2$   &  $\delta$  &    $V_{c}$  &  $V_a$ \\ 
%Data    & $\!\!\!\!\!\!\!\!\!\!(\unit[]{kpc^2Myr^{-1}})\!\!\!\!\!\!\!\!\!\!$  &    &  $\!\!\!\!\!\!\!\!\!(\unit[]{km\,s^{-1})}\!\!\!\!\!\!\!\!\!$ & $\!\!\!\!\!\!\!\!\!\!\!(\unit[]{km\,s^{-1}})\!\!\!\!\!\!\!\!\!\!\!$ \\\hline
%& \multicolumn{4}{c}{} \\ 
%II-F    & $8.6^{+0.2}_{-0.2}$  & $\!\!\!\!\!0.239^{+0.005}_{-0.007}\!\!\!\!\!$ & $\cdots$ & $69^{+2}_{-2}$ \vspace{0.1cm}\\
%III-F   & $0.41^{+0.04}_{-0.07}$  & $0.86^{+0.04}_{-0.03}$ &  $18.7^{+0.4}_{-0.3}$ & $35^{+2}_{-2}$ \vspace{0.1cm}\\
%\hline
%\end{tabular}
%\end{table}

\begin{table}[t]
\caption{Best-fit model parameters on B/C data: 1D versus 2D DM ($L=4$~kpc).}
\label{tab:1Dvs2D}
\centering
\begin{tabular}{lcccccc} \hline\hline
Model    & $\!\!K_0^{\rm best}\times 10^2\!\!$     & $\delta^{\rm best}$ &  $V_c^{\rm best}$ &  $V_a^{\rm best}$  & $\!\!\!\!\chi^2$/d.o.f$\!\!\!\!$   \\
Data        & $\!\!\!\!\!\!$(kpc$^2\,$Myr$^{-1}$)$\!\!\!\!\!\!$ &        &  (km$\;$s$^{-1}$)  & (km$\;$s$^{-1}$)   & \\\hline
& \multicolumn{5}{c}{} \\ 
1D~~II-F $\!\!\!\!\!\!$& 9.74 & 0.23 & \dots& 73.1 &  4.68  \vspace{0.05cm}\\  %chi2=132.499 (31)
2D~~II-F  $\!\!\!\!\!\!$& 8.56 & 0.24 & \dots & 68.6 &  4.67  \vspace{0.15cm}\\  %chi2=130.789 (28)
1D~~III-F$\!\!\!\!\!\!$& 0.48 & 0.86 & 18.9 & 38.0 &  1.47  \vspace{0.05cm}\\  %chi2=39.7026 (31)
2D~~III-F $\!\!\!\!\!\!$& 0.42 & 0.86 & 18.7 & 35.5 &  1.46  \vspace{0.05cm}\\  %chi2=39.2886 (27)
\hline
\end{tabular}
\end{table}
%   

%%%%%%%%%%%%%%%%%%%%%%%%%%%%%%%%%%%%%%%%%%%%%%%%%%%%%%%%%%%%%%%%%%%%%%
%%%%%%%%%%%%%%%%%%%%%%%%%%%%%%%%%%%%%%%%%%%%%%%%%%%%%%%%%%%%%%%%%%%%%%
\section{Results for radioactive species (free halo size $L$)\label{sec:1D_rad}}

We now attempt to lift the degeneracy between the halo size and the normalisation
of the diffusion coefficient, using radioactive nuclei.
The questions that we wish to address are the following: i) With existing data,
how large are the uncertainties in $L$ for a given model? ii)
Do radioactive nuclei provide different answers for models
with different $\delta$? iii) Is the mean value (and uncertainty) for $L$
obtained from a given isotopic/elemental ratio consistent with or stronger constrained
than that obtained from another measured isotopic/elemental ratio?
iv) How does the presence of a local underdense bubble (modelled as a hole of radius $r_h$,
see Sect.~\ref{sec:rad}) affect the conclusions?

Until now, almost all studies have focused on the isotopic ratios of $^{10}$Be/$^{9}$Be, $^{26}$Al/$^{27}$Al,
$^{36}$Cl/Cl, and  $^{54}$Mn/Mn. An alternative, discussed in \citet{1998ApJ...506..335W}, is to
consider the Be/B, Al/Mg, Cl/Ar, and Mn/Fe ratios.  The advantage of considering these elemental ratios is
that they are easier to measure than isotopic ratios, and thus provide a wider energy range to which
we can fit the data. Taking ratios such as Be/B maximises the effect of radioactive
decay, since the numerator represents the decaying nucleus and the denominator the decayed nucleus.
However, the radioactive contribution is only a fraction of the elemental flux,
and HEAO-3 data were found to be less constraining that the isotopic ratios in \citet{1998ApJ...506..335W}.

Below, we consider and compare the constraints from both the isotopic ratios and
the elemental ratios. The data used are described in Appendix~\ref{app:data_rad}.
We discard $^{54}$Mn because it suffers more uncertainties than the others in the calculation
(and also experimentally) due to the electron capture decay channel. 
The free parameters for which we seek the PDF are the four transport parameters
$\{K_0,\,\delta,\,V_c,\,V_a\}$, plus one $\{L\}$ or two geometrical parameters $\{L,\, r_h\}$,
depending on the configuration considered. The main results of this section are thus in
identifying the PDF of $L$ for the standard DM, and the PDFs of both $L$ and $r_h$ for
the modified DM.

             %%#######################################%%
 \subsection{PDFs of $L$ and $r_h$ using isotopic measurements}
We start with a simultaneous fit to B/C and $^{10}$Be/$^{9}$Be,
for both model III (diffusion/convection/reacceleration),
and model II (diffusion/reacceleration), the latter being frequently
used in the literature.

 \subsubsection{Simultaneous fit to B/C and $^{10}$Be/$^{9}$Be}
\begin{figure*}[!th]
	\includegraphics[width = 0.667\textwidth]{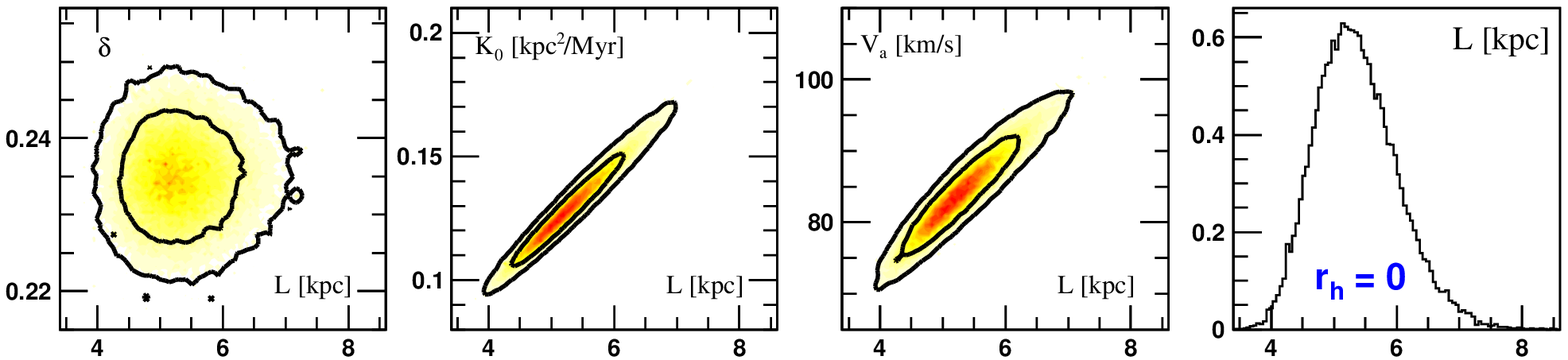}
	\vspace{0.25cm}\\
	\begin{minipage}{\textwidth}
    \hrule width 12.75cm	\vspace{0.45cm}
	\end{minipage}\\
	\includegraphics[width = 0.835\textwidth]{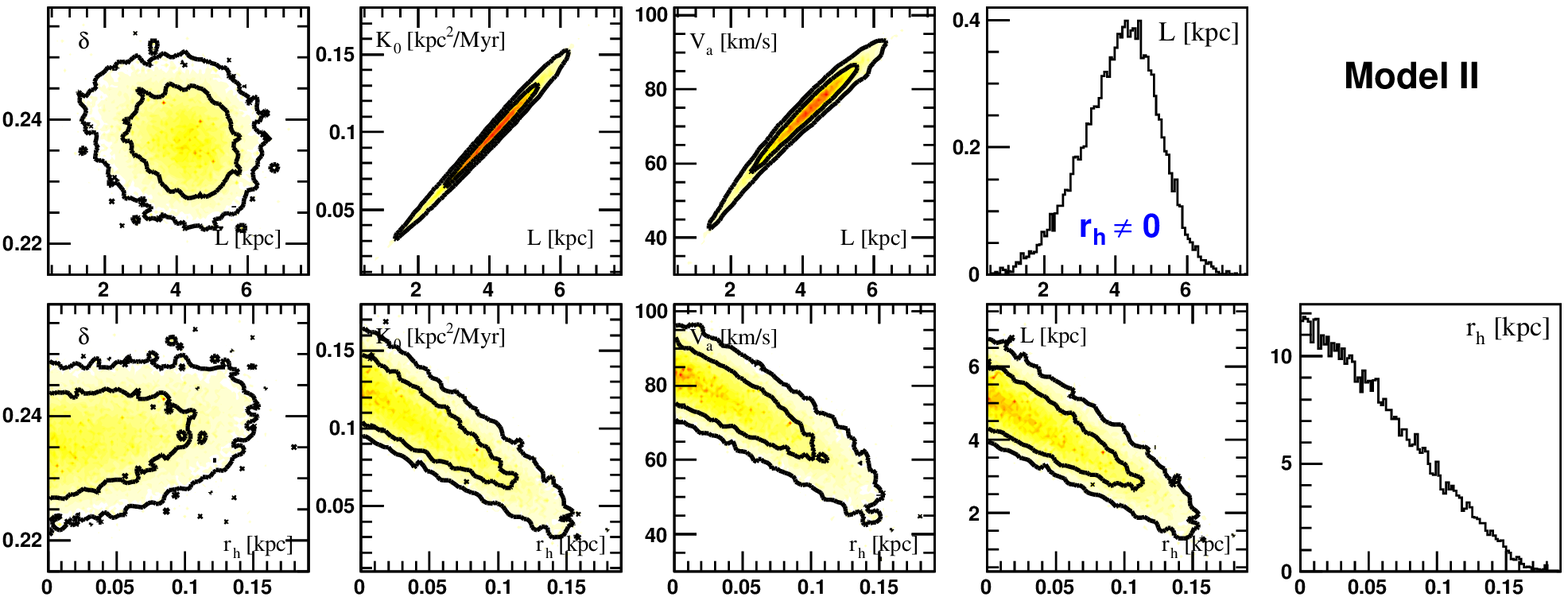}
	\caption{Model II (diffusion/reacceleration): marginalised posterior PDF of the diffusive halo size $L$ (right panels of the first and second row) and the local bubble radius $r_h$ (right panel of the last row) for the standard ($r_h=0$, first row) and the modified DM ($r_h \neq 0$, second and last row), as constrained by using B/C and $^{10}$Be/$^9$Be data. The correlations between the geometrical parameters $L$ and $r_h$ and the transport parameters $\delta$, $K_0$, and $V_a$, are shown in the 2D histograms. The colour code corresponds to the regions of increasing probability (from paler to darker shades), and the two contours (smoothed) delimit regions containing respectively 68\% and 95\% (inner and outer contour) of the PDF.}
	\label{fig:ModelIIF_Lrh}
\end{figure*}

\begin{figure*}[!t]
	\includegraphics[width = 0.835\textwidth]{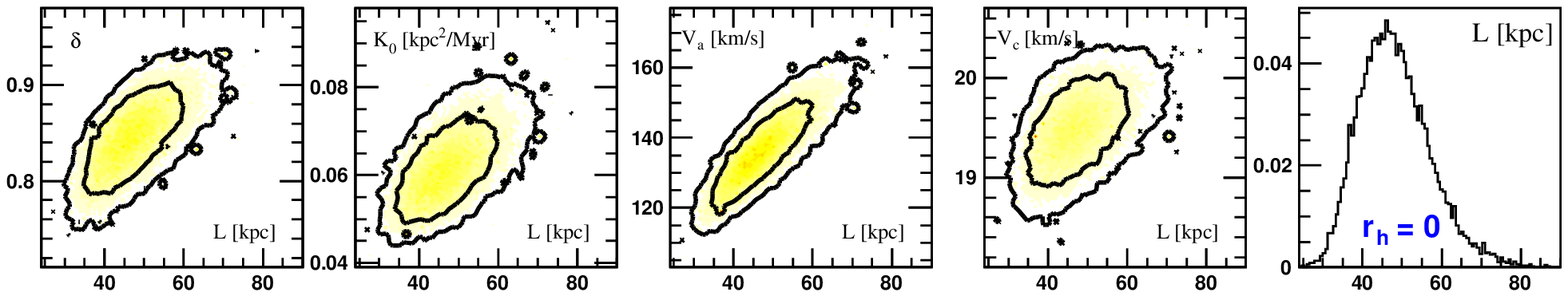}
	\vspace{0.25cm}\\
	\begin{minipage}{\textwidth}
    \hrule width 15.75cm	\vspace{0.45cm}
	\end{minipage}\\
	\includegraphics[width = \textwidth]{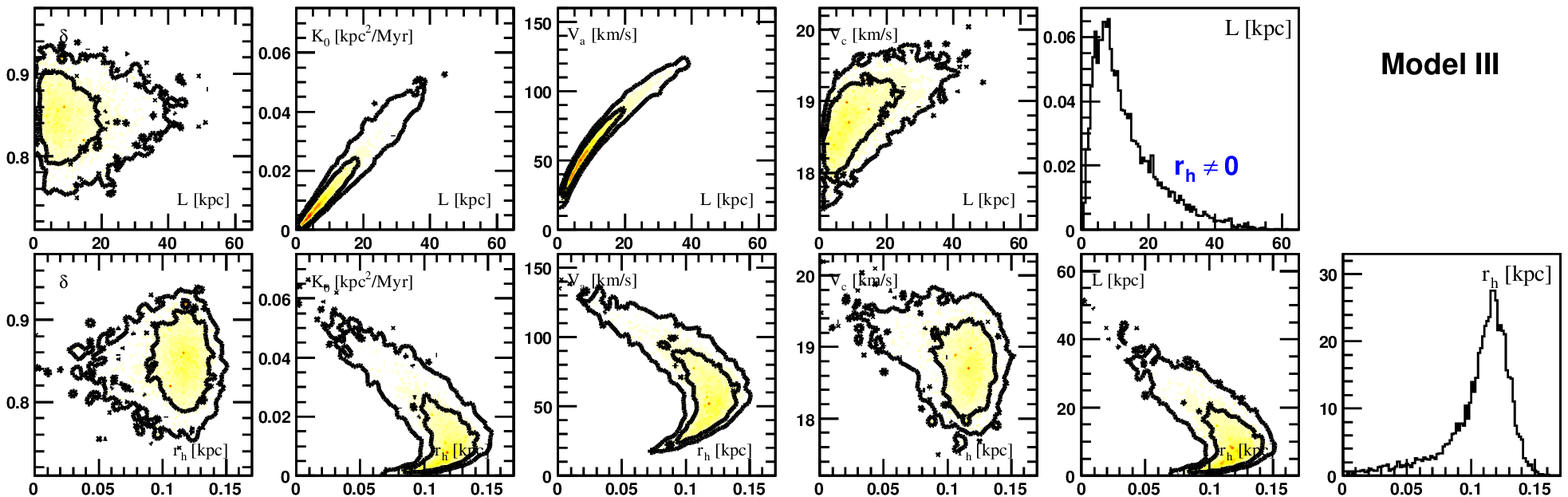}
	\caption{Model III (diffusion/convection/reacceleration): same as in Fig.~\ref{fig:ModelIIF_Lrh}. The transport parameters are now $\delta$, $K_0$, $V_a$, and $V_c$, with the geometrical parameters $L$ and $r_h$.}
	\label{fig:ModelIIIF_Lrh}
\end{figure*}

 The marginalised posterior PDFs of $L$ and $r_h$ and the correlations between these new free parameters and
the propagation parameters of models~II and III are given in the Figs.~\ref{fig:ModelIIF_Lrh} and~\ref{fig:ModelIIIF_Lrh}, respectively.
The most probable values of the parameters are gathered in Table~\ref{tab:rad_results}.
\begin{table}[!tb]
\caption{Most probable values for models~II and III for the free parameters of
the local bubble radius $r_h$ and/or the Galactic halo size $L$
(constrained by B/C and $^{10}$Be/$^9$Be data).}
\label{tab:rad_results}\centering
\begin{tabular}{lcccccc} \hline\hline
        & $K_{0} \times 10^2$   &  $\delta$  &    $V_{c}$  &  $V_a$ & $L$ & $r_h$ \\ 
        & $\!\!\!\!\!\!\!\!\!\!(\unit[]{kpc^2Myr^{-1}})\!\!\!\!\!\!\!\!\!\!$  &    &  $\!\!\!\!\!\!\!\!\!\!\!\!(\unit[]{km\,s^{-1})}\!\!\!\!\!\!\!\!\!\!\!\!$ & $\!\!\!\!\!\!\!\!\!\!\!(\unit[]{km\,s^{-1}})\!\!\!\!\!\!\!\!\!\!\!$ & $(\unit[]{kpc})$ & $(\unit[]{pc})$\\\hline
& \multicolumn{4}{c}{} \\ 
II  & $8.6^{+0.2}_{-0.2}$  & $\!\!\!\!\!\!\!\!0.239^{+0.005}_{-0.007}\!\!\!\!\!\!\!\!$ & $\cdots$ & $69^{+2}_{-2}$ & [4] & $\cdots$ \vspace{0.1cm}\\
II  & $13^{+2}_{-2}$  & $\!\!\!\!\!\!\!\!0.234^{+0.006}_{-0.005}\!\!\!\!\!\!\!\!$ & $\cdots$ & $84^{+5}_{-6}$ & $\!\!\!\!\!\!\!5.2^{+0.7}_{-0.6}\!\!\!\!\!\!\!$ & $\cdots$ \vspace{0.1cm}\\
II  & $11^{+2}_{-3}$  & $\!\!\!\!\!\!\!\!0.235^{+0.008}_{-0.004}\!\!\!\!\!\!\!\!$ & $\cdots$ & $77^{+8}_{-11}$ & $4^{+1}_{-1}$ & $\!\!\!3^{+70}_{-3}\!\!\!$ \vspace{0.1cm}\\
III$\!\!\!$ & $0.41^{+0.04}_{-0.07}$  & $0.86^{+0.04}_{-0.03}$ &  $\!\!\!18.7^{+0.4}_{-0.3}\!\!\!$ & $35^{+2}_{-2}$ & [4] & $\cdots$ \vspace{0.1cm}\\
III$\!\!\!$ & $6.1^{+0.8}_{-0.8}$  & $0.86^{+0.03}_{-0.05}$ &  $\!\!\!19.4^{+0.4}_{-0.3}\!\!\!$ & $135^{+11}_{-10}$ & $\!\!\!\!\!46^{+9}_{-8}\!\!\!\!\!$ & $\cdots$ \vspace{0.1cm}\\
III$\!\!\!$ & $0.8^{+1}_{-0.7}$  & $0.86^{+0.03}_{-0.04}$ &  $\!\!\!18.7^{+0.5}_{-0.4}\!\!\!$ & $55^{+31}_{-21}$ & $8^{+8}_{-7}$ & $\!\!\!\!\!\!\!\!120^{+20}_{-20}\!\!\!\!\!\!\!\!$ \vspace{0.1cm}\\
\hline
\end{tabular}
\end{table}

For all configurations, the diffusion slope $\delta$ and the Galactic wind $V_c$ are unaffected by the addition of the free parameters $L$ and $r_h$. The B/C fit is degenerate in $K_0/L$ and $V_a/\sqrt{K}$,
so that the values of $K_0$ and $V_a$ vary as $L$ varies. For model~III, their evolution follows the relations given in Fig.~\ref{fig:BCparam_vs_L}. This implies that there is a positive correlation between $K_0$ and $V_a$, and $K_0$ and $L$, as seen
from Figs.~\ref{fig:ModelIIF_Lrh} and \ref{fig:ModelIIIF_Lrh}.
The uncertainty in the diffusive halo size $L$ is smaller for II than for III. This is a consequence of the inclusion of the constant wind, which decreases the resolution on $K_0$ from 2\% (Model~II) to 10\% (Model~III)|see e.g., Table~\ref{tab:DM_results} or \ref{tab:rad_results}|hence broadening the distribution of $L$. 

Below, the results for the standard DM|for which $r_h$ is set to be 0|and those
for the modified DM|for which $r_h$ is left as an additional free parameter|are
discussed separately.  This allows us to emphasise the impact of $r_h$ on the other parameters, which
is different for models~II and~III.
\paragraph{Standard DM ($r_h=0$):} the parameter $L$ is constrained to be between $4.6$ and $\unit[5.9]{kpc}$ for model~II, having a most probable value at \unit[5.2]{kpc}|a result compatible with other studies \citep{2001ICRC....5.1836M}|the posterior PDF of $L$ extends from 25 to \unit[85]{kpc} for model~III (most probable value at \unit[46]{kpc}). In terms of statistics, the best-fit model is stil model~III, for which the $\chi^2$/d.o.f is 1.41. 

\paragraph{Modified DM ($r_h\neq0$):} 
the presence of a local bubble results in an exponential attenuation of the local radioactive flux,
see Sect.~\ref{sec:rad} and Eq.~(\ref{eq:rad_damping}). We thus expect to have a different best-fit parameter for $L$ in that case.
The resulting posterior PDFs of $L$ and $r_h$ and the correlations to the propagation parameters for this modified DM are given in Figs.~\ref{fig:ModelIIF_Lrh} and~\ref{fig:ModelIIIF_Lrh} for models~II and III respectively. The most probable values are gathered in Table~\ref{tab:rad_results} (third and last lines).

As expected, the local bubble radius $r_h$ is negatively correlated with the Galactic halo size $L$. The effect is more striking for model~III, where the favoured range for $L$ extends from 1 to \unit[50]{kpc}. The most probable value is \unit[$8^{+8}_{-7}$]{kpc} for a local bubble radius $r_h = \unit[120^{+20}_{-20}]{pc}$. The $\chi^2$/d.o.f of this configuration is 1.28, instead of 1.41 for the standard DM. The improvement to the fit is statistically significant according to the Fisher criterion. 

The situation for model~II is different. The halo size $L$ is already small for the standard configuration $r_h=0$. Adding the local bubble radius $r_h$ to the fit decreases the most probable value of $L$ only slightly to $\unit[4^{+1}_{-1}]{kpc}$ and the measured value of $r_h$ is compatible with \unit[0]{pc}. In addition, the $\chi^2$/d.o.f is 3.69 and hence poorer than for the configuration without the local bubble feature. In this model
(diffusion/reacceleration, no convection), a local underdensity is not supported.

\subsubsection{Results and comparison with fits to $^{26}$Al/$^{27}$Al and $^{36}$Cl/Cl}
\begin{figure}[t]
	\hspace{0.15cm}\includegraphics[width = 0.235\textwidth]{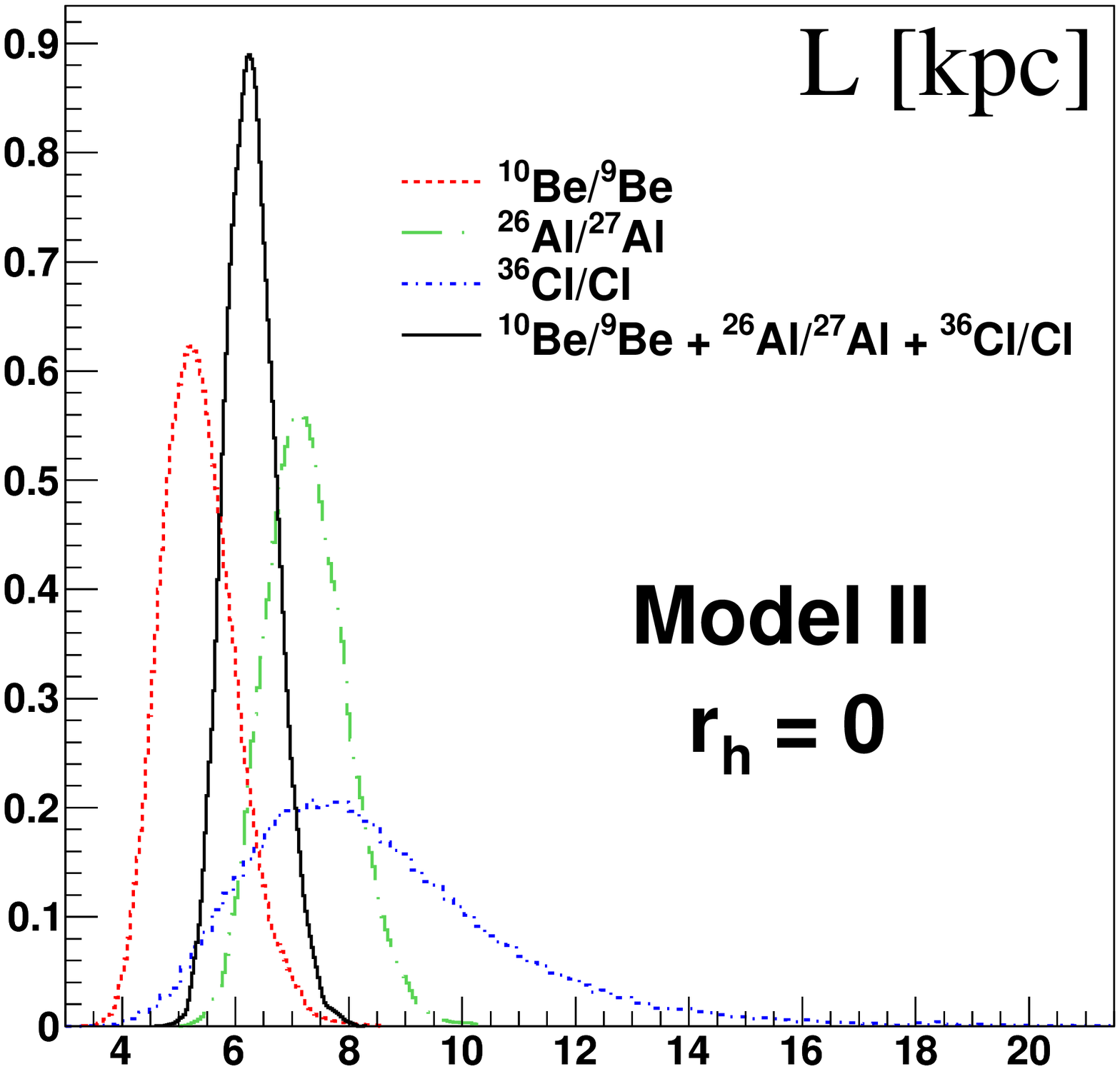}\\
	\includegraphics[width = 0.5\textwidth]{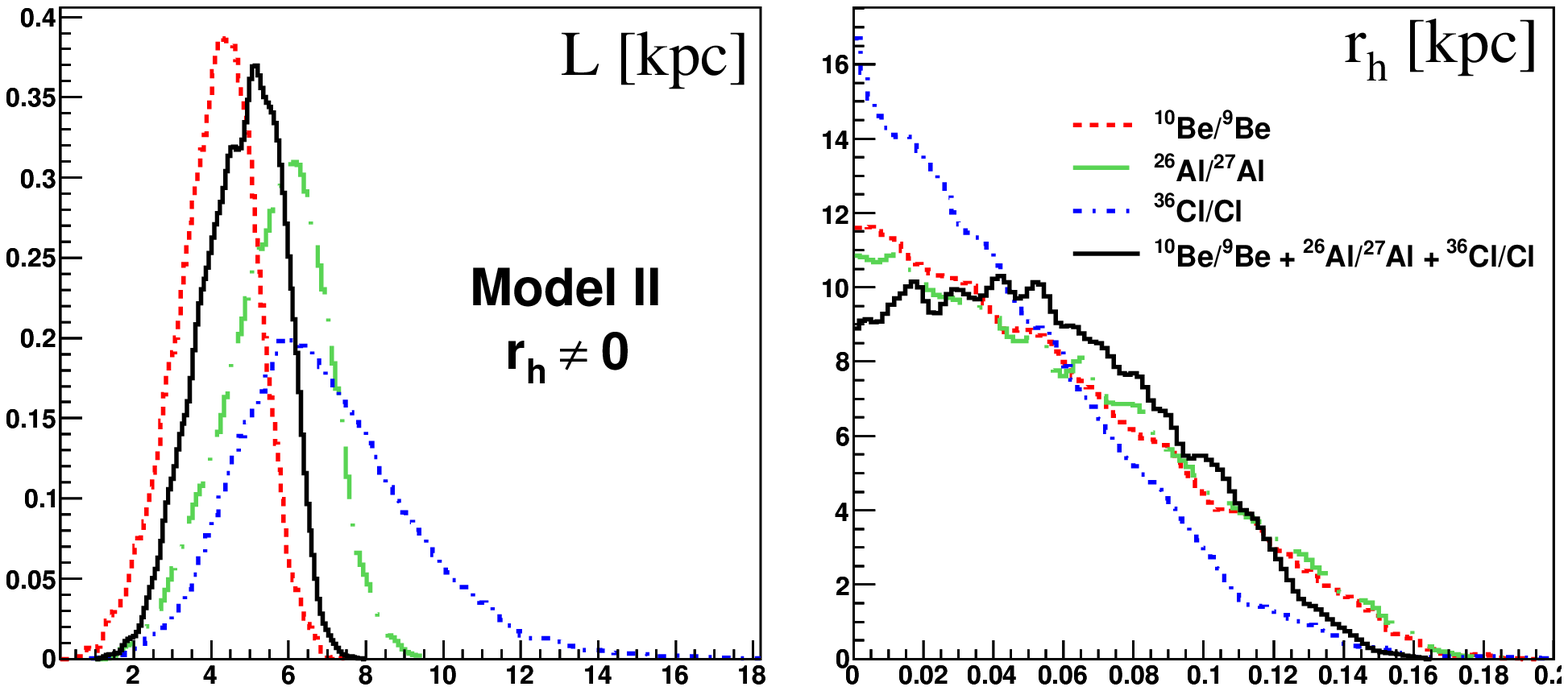}
	\caption{Model II: marginalised posterior PDFs of the Galactic geometry parameters for the standard DM ($r_h=0$, top panel) and for the modified DM ($r_h\neq0$, bottom panels). The four curves result from the combined analysis of B/C plus isotopic ratios of radioactive species: B/C$+^{10}$Be/$^9$Be (red dotted line), B/C$+^{26}$Al/$^{27}$Al (green long dashed-dotted line), and B/C$+^{36}$Cl/Cl (blue dashed-dotted line). The black solid curve represents the extracted PDF resulting from a simultaneous fit of B/C plus all three isotopic ratios.
All PDFs are smoothed.}
	\label{fig:ModelIIF_L_Lrh_isotope}
\end{figure}

\begin{figure}[t]
	\hspace{0.15cm}\includegraphics[width = 0.235\textwidth]{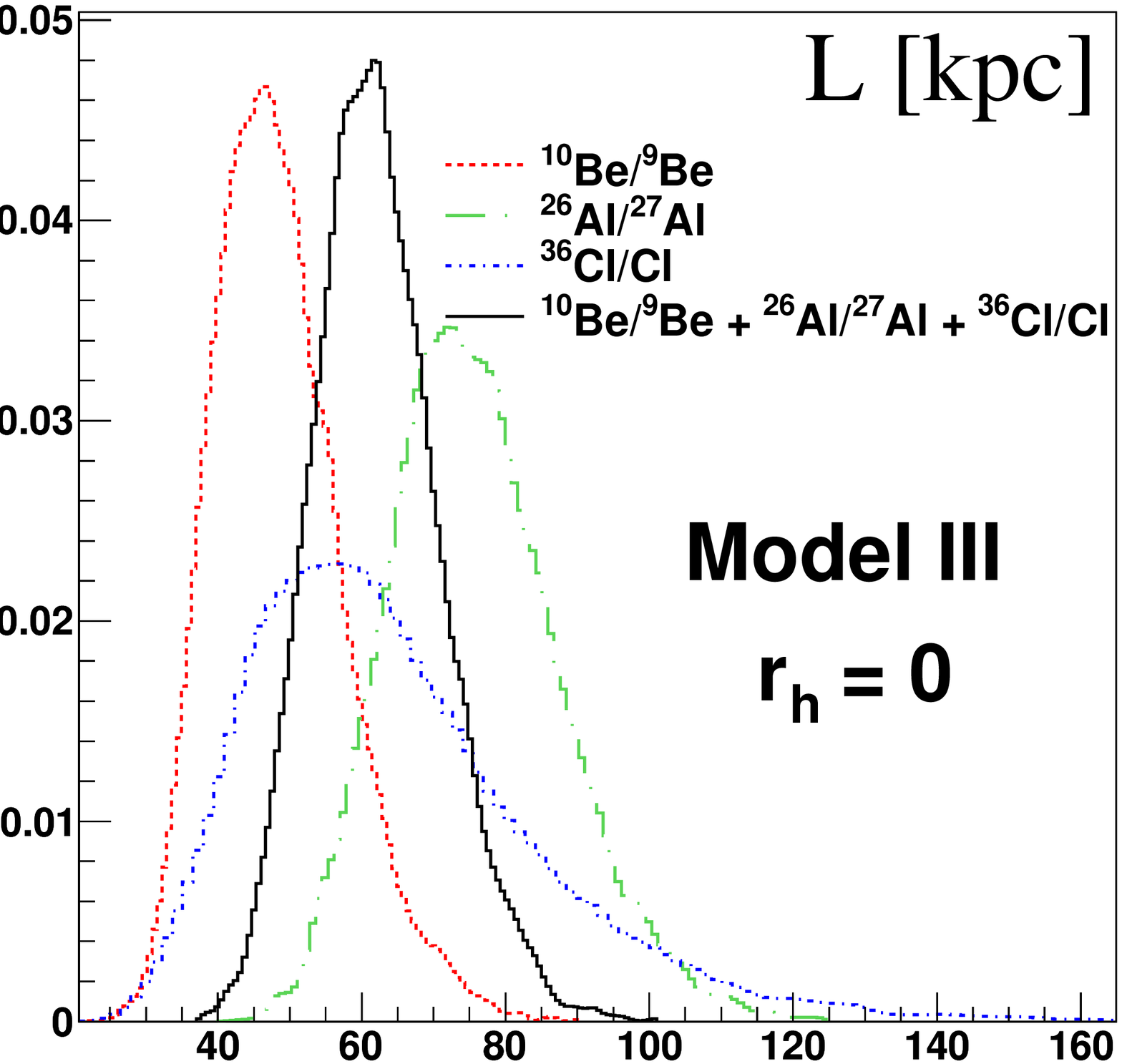}\\
	\includegraphics[width = 0.5\textwidth]{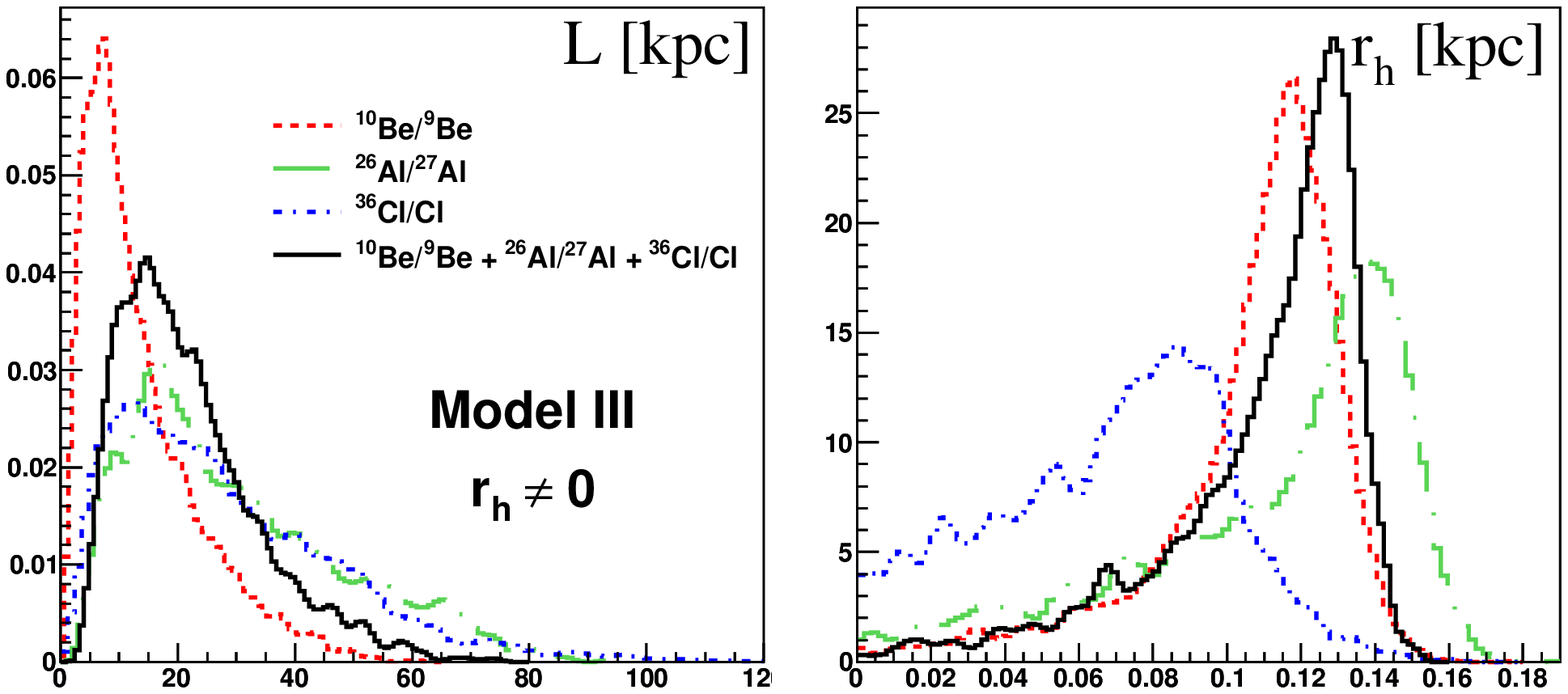}
	\caption{Same as in Fig.~\ref{fig:ModelIIF_L_Lrh_isotope}, but for model~III.}
	\label{fig:ModelIIIF_L_Lrh_isotope}
\end{figure}

We repeat the analysis for the remaining isotopic ratios.
The resulting marginalised posterior PDFs of the Galactic halo size $L$ and the local underdensity $r_h$ are given in Figs.~\ref{fig:ModelIIF_L_Lrh_isotope} and \ref{fig:ModelIIIF_L_Lrh_isotope} for models~II and III, respectively. The correlation plots with the transport parameters are similar to those of
Figs.~\ref{fig:ModelIIF_Lrh} and~\ref{fig:ModelIIIF_Lrh} and are not repeated.

\paragraph{Standard DM ($r_h=0$):} 
as for the $^{10}$Be/$^9$Be ratio (red-dotted line), $L$ is well constrained in model~II at small values for
the $^{26}$Al/$^{27}$Al (green-long dashed-dotted line), and $^{36}$Cl/Cl (blue dashed-dotted line) ratios,
covering slightly different but consistent ranges from 4 to \unit[14]{kpc}. The width of the estimated PDFs
increases when moving from the $^{10}$Be/$^9$Be ratio to the $^{36}$Cl/Cl ratio, due to the decreasing
accuracy of the data. In the same way, the adjustment to the data becomes poorer, as expressed by the increase
in $\chi^2$/d.o.f from 3.59 to 4.09. Used alone, the radioactive ratio $^{10}$Be/$^9$Be constrains 
the most precisely the halo size $L$, but the constraints obtained with the other radioactive ratios are
completely compatible within the 2$\sigma$ range. The most likely value of $L$ is ascertained when the three
radioactive ratios are fitted simultaneously (black solid line).

The best-fit model is model~III, where the overall covered halo size range extends from 20 to
\unit[140]{kpc}. The most probable value found for $L$ with 68\% confidence level (CL) errors is
$\unit[62^{+7}_{-10}]{kpc}$.

\paragraph{Modified DM ($r_h\neq0$):} 
the resulting marginalised posterior PDFs of $L$ are shown in Figs.~\ref{fig:ModelIIF_L_Lrh_isotope} and \ref{fig:ModelIIIF_L_Lrh_isotope} (lower left) for models~II and III, respectively. Again, the extracted PDFs for all radioactive ratios are completely compatible for both models.
As described above, the decrease in $L$ is more pronounced for model~III than for model~II. This decrease can be observed for all radioactive ratios, independently of the model chosen. 

The resulting marginalised posterior PDFs of $r_h$ are given in Figs.~\ref{fig:ModelIIF_L_Lrh_isotope} 
and \ref{fig:ModelIIIF_L_Lrh_isotope} (lower right) for models~II and III, respectively. 
The addition of an underdensity in the local interstellar medium is preferred by the data in the best-fit model~III, but it is disfavoured in model~II. The most probable values for $r_h$ range from \unit[90]{pc} for the $^{36}$Cl/Cl ratio to \unit[140]{pc} for the the $^{26}$Al/$^{27}$Al ratio, and the overall fit points to a most probable radius of \unit[$130^{+10}_{-20}$]{pc}.

These results confirm and extend the slightly different analysis of \citet{2002A&A...381..539D}, who found that for model~III, the best-fit values
for $r_h$ was $\sim 80$~pc (see also Appendix~\ref{sec:localbubble}).

\subsubsection{Envelopes of 68\% CL\label{sec:CL}}
\begin{figure*}[!t]
	\includegraphics[width = \columnwidth]{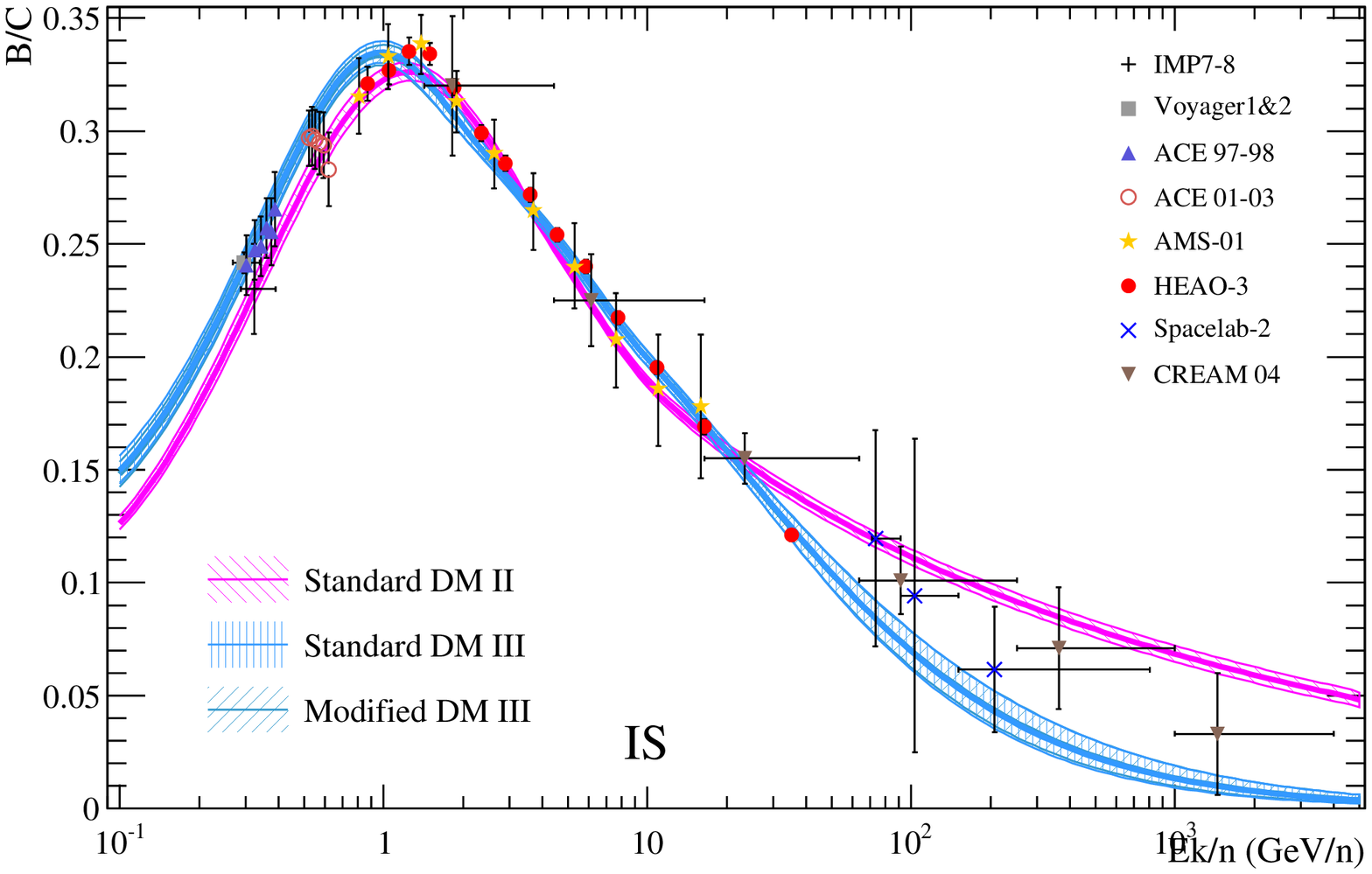}
	\includegraphics[width = \columnwidth]{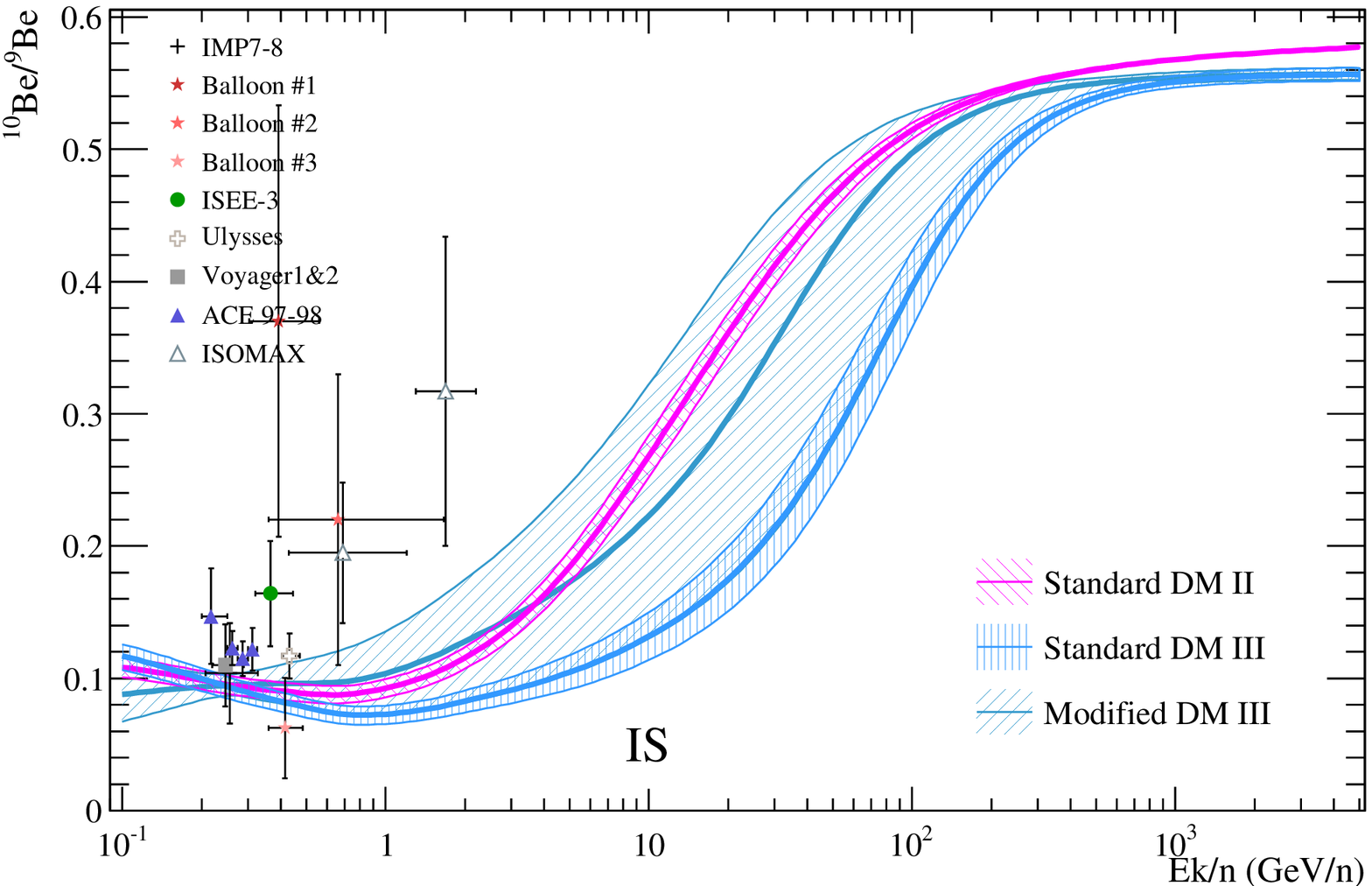}\\
	\includegraphics[width = \columnwidth]{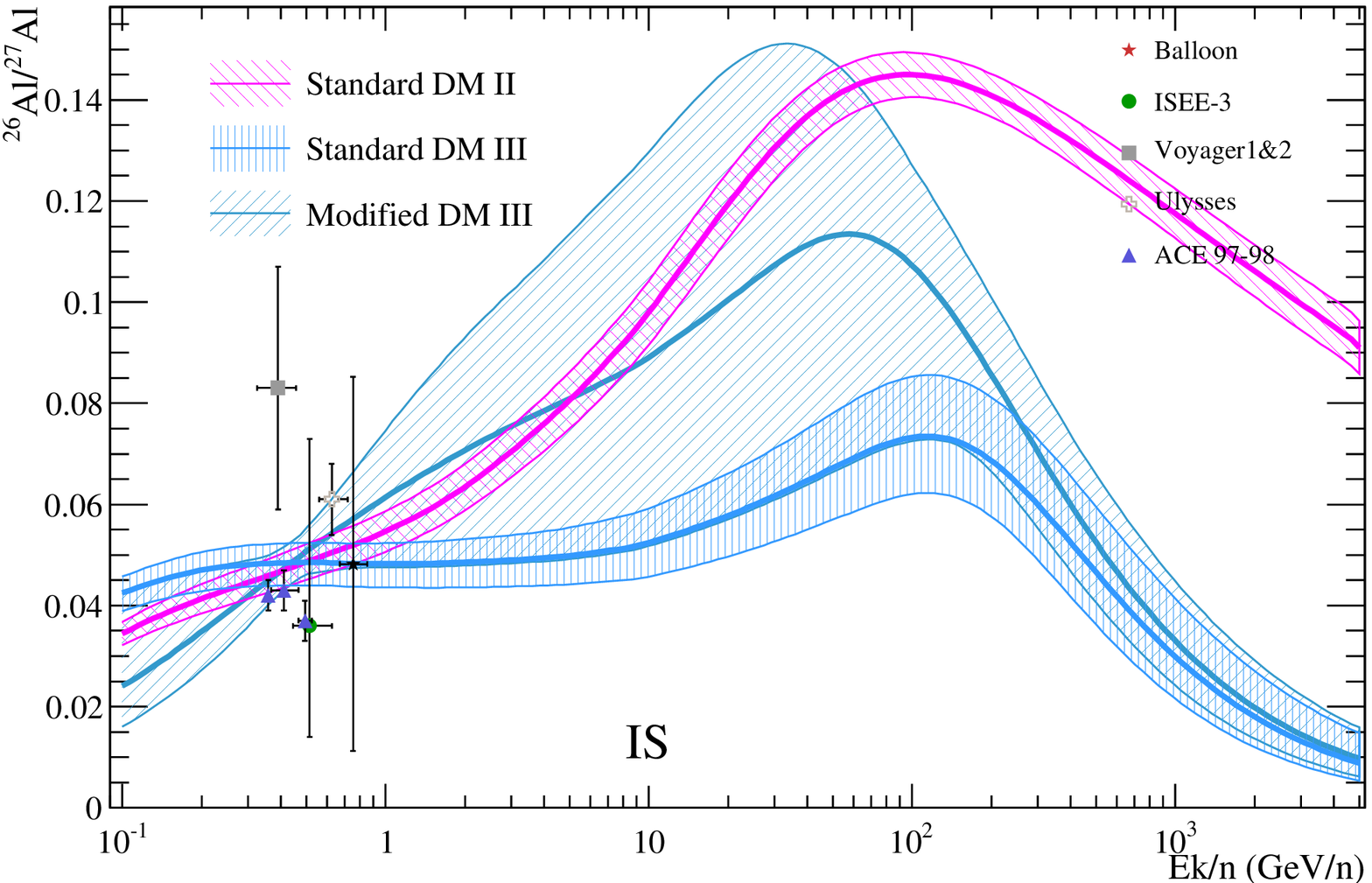}
	\includegraphics[width = \columnwidth]{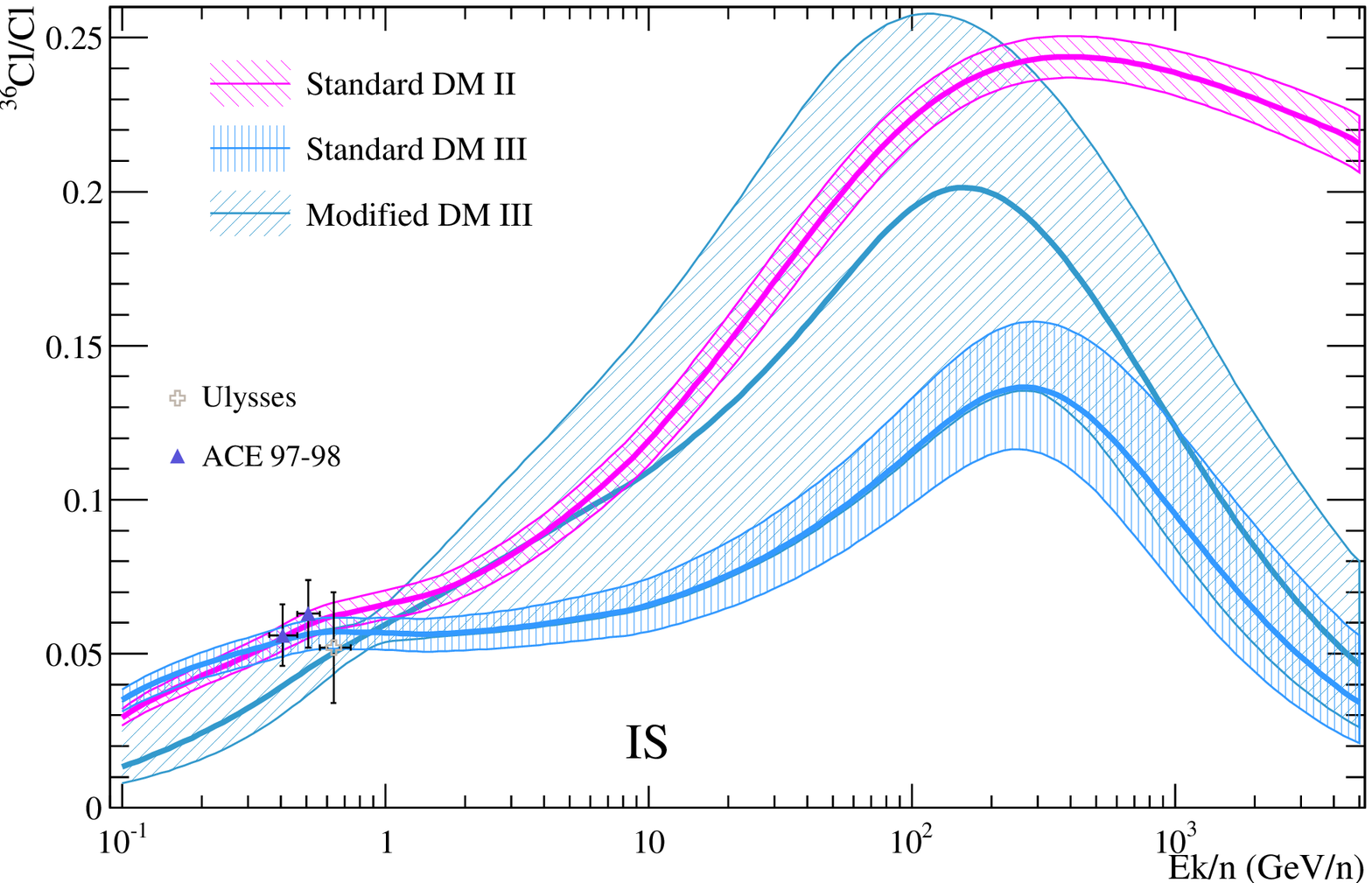}
   \caption{Shown are the envelopes of 68\% CL (shaded areas) and best-fit (thick lines) ratios for the standard
   DM~II ($r_h = 0$, red) and for model~III (standard and modified DM, blue) in the 1D geometry (based on the B/C + $^{10}$Be/$^9$Be + 
   $^{26}$Al/$^{27}$Al + $^{36}$Cl/Cl constraint). All quantities are IS. The data are demodulated
   using the approximate procedure $E_k^{IS} =  E_k^{\rm TOA} + \Phi$.}
	\label{fig:envelopes}
\end{figure*}

Confidence contours (for any combination of the CR fluxes) corresponding to given confidence levels (CL) in
the $\chi^2$ distribution can be drawn, as detailed in Appendix~A and Sect.~5.1.4 of Paper~I. From the MCMC
calculation based on the B/C + $^{10}$Be/$^9$Be + $^{26}$Al/$^{27}$Al + $^{36}$Cl/Cl constraint, we select
all sets of parameters for which the $\chi^2$ meets the 68\% confidence level criterion. For each
set of these parameters, we calculate the B/C and the three isotopic ratios. We store for each energy the minimum and maximum value of the ratio. The
corresponding contours (along with the best-fit ratio) for models~II (standard DM, red) and III (standard and modified DM, blue) are drawn in
Fig.~\ref{fig:envelopes}.
To ease the comparison with the data, all results correspond to IS quantities
(the approximation made in the demodulation procedure, see Paper I, is negligible
with respect to the experimental error bars).

We see that the present data already constrain very well the various ratios for the standard DM. 
The difference between the results of models~II and III are more pronounced at high energy
(effect of $\delta$), as seen from the B/C ratio beyond 10~GeV/n.
All contours are pinched around 10~GeV/n, which is a consequence of the energy chosen to
renormalise the flux to the data in the propagation code. In principle, the source abundance
of each species may be set as an additional free parameter in the fit (Paper~I), but at the cost of
the computing time. The three isotopic ratios ($^{10}$Be/$^9$Be$^{26}$, Al/$^{27}$Al,
and $^{36}$Cl/Cl) provide a fair match to the data for all models, considering the large scatter and
possible inconsistencies between the results quoted by various experiments. In particular,
for $^{10}$Be/$^{9}$Be, new data are necessary to confirm the high value of the ratio measured
at $\sim$~GeV/n energy.

The envelope for the modified DM is quite large at high energy, because the uncertainty in
$r_h$ is responsible for a larger scatter in the other parameters. The two standard DM
contain non-overlapping envelopes beyond GeV/n energies. This means that to disentangle the models,
having measurements of the above isotopic ratios in the $1-10$~GeV/n may be more important
than just having more and higher quality data at low energy.

\subsubsection{General dependence of $L$ with $\delta$ (for $r_h=0$)}

To investigate the difference in the results obtained from models II and III, we fit B/C and the
three isotopic ratios for different values of $\delta$ (a similar trend with $L$ is obtained if
just one isotopic ratio is selected). The analysis relies on the Minuit minimisation routine to
quickly find the best-fit values, as described in \citet{2010A&A...xxx..xxxM}. %
\begin{figure*}[t]
\centering
\includegraphics[width = \columnwidth]{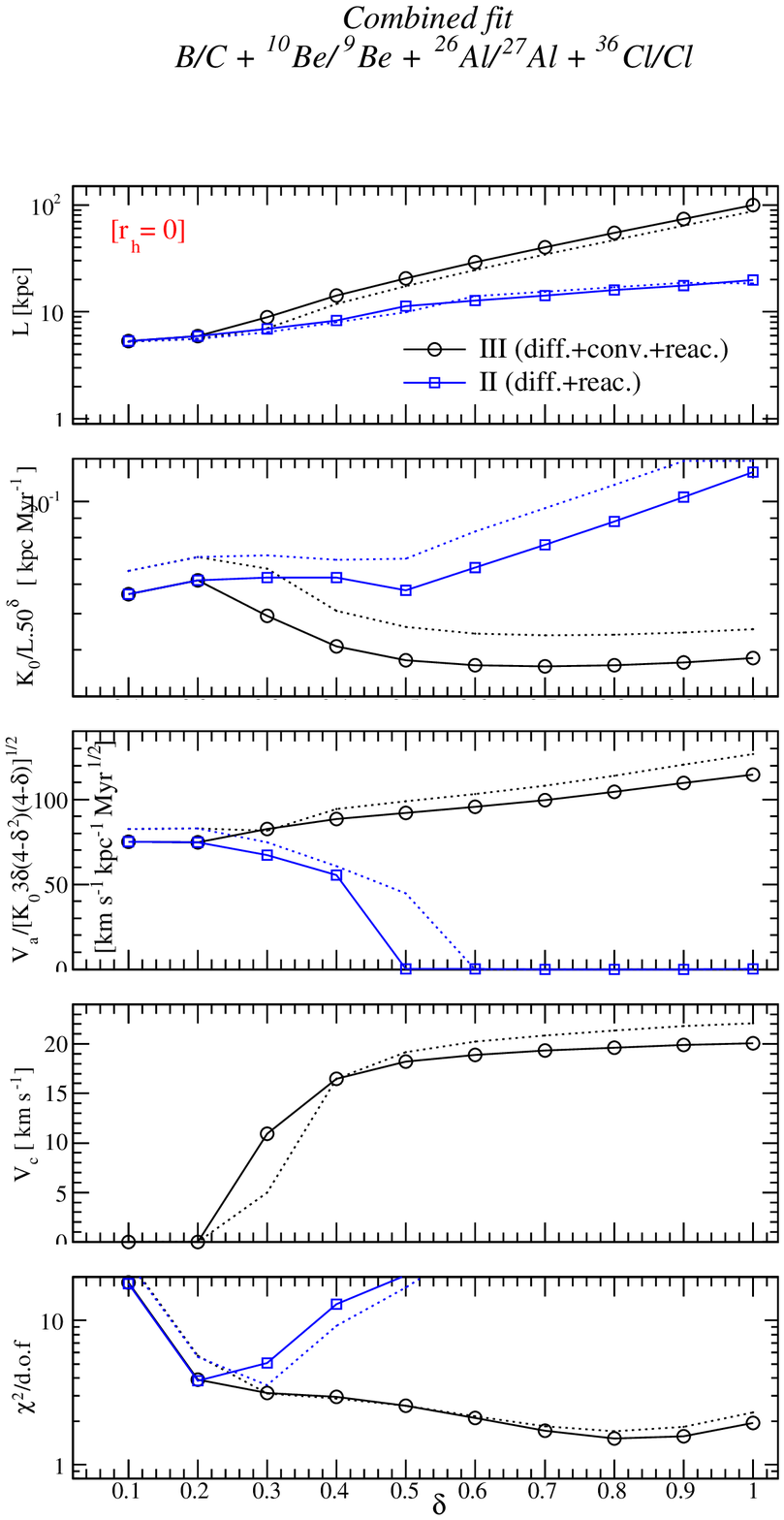}
\includegraphics[width = \columnwidth]{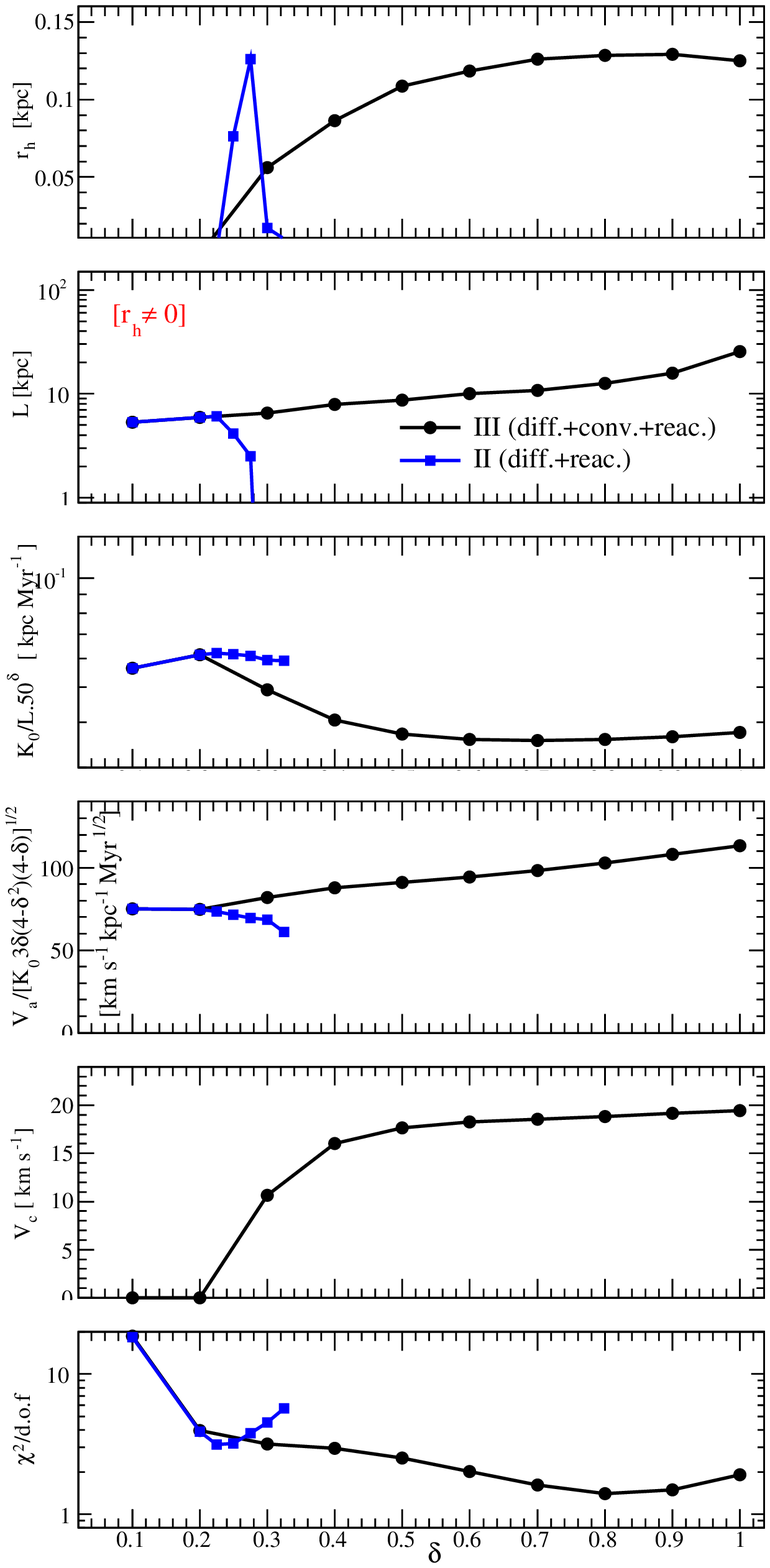}
\caption{Left panel: standard DM model ($r_h=0$)|thin-dotted lines
are derived using the GAL09 instead of the W03 fragmentation cross-sections. Right panel: modified DM model ($r_h\neq0$). 
For both panels, shown are the best-fit parameters on B/C + $^{10}$Be/$^9$Be + $^{26}$Al/$^{27}$Al + $^{36}$Cl/Cl
data, as a function of the diffusion slope $\delta$. The latter is varied between 0.1 and 1.0 for model
II (blue lines, open and filled squares) and model III (black lines, open and filled circles).
From top to bottom, $L$, $K_0/L$, $V_a/\sqrt{K_0}$, and $V_c$ as a function of $\delta$ are shown.
The bottom panel shows the best $\chi^2/$d.o.f for each $\delta$.}
\label{fig:paramL_vs_delta}
\end{figure*}
The evolution of the parameters with $\delta$ is shown on the left side of Fig.~\ref{fig:paramL_vs_delta}.
The bottom panel shows the evolution of $\chi^2_{\rm min}/$d.o.f, where we recover that the
best-fit $\delta$ for model II (dashed-blue line) lies around $\delta\approx 0.2$, whereas that for model III
(solid-black line) lies around $\delta\approx 0.8$. As already underlined, the contribution to the $\chi^2$ value
is dominated by the B/C contribution because as discussed in Appendix~\ref{sec:MCMC_combi}, the
values of transport parameters that reproduce the B/C ratio are expected to remain within a narrow range.
This explains what is observed in the various panels showing these combinations. For model~II, we emphasise
that for $\delta\gtrsim0.5$, the best-fit value for $V_a$ is zero (Model~II becomes a pure diffusion model).

The most important result,  given in the top panel, is for $L$ as a function
of $\delta$, where any uncertainty in the determination of $\delta$ translates
into an uncertainty in the determination of $L$. When a Galactic wind is considered (Model III,
black-solid line), the correlation between $L$ and $\delta$ is stronger than
for model II (no wind). There is no straightforward explanation of this dependence.
The flux of the radioactive isotope can be shown to be $N^{\rm rad}(0)\approx hq/\sqrt{K\gamma\tau_0}$
\citep[e.g.,][]{2006astro.ph.12714M}. Since secondary fluxes should match the data regardless of the value for $\delta$,
this implies that the ratio $^{10}$Be/$^9$Be depends only on $\sqrt{K\gamma\tau_0}$.
At the same time, to ensure that the secondary-to-primary ratio is constant, we must maintain
a constant
$L/K$. The difficulty is that the former quantity is a constant at low rigidity where
the isotopic ratio is measured, whereas the latter quantity should remain as close a possible to the B/C
data over the whole energy range. Hence, all we can say is that the variation in $L$ with $\delta$
is related to the variation in $K_0/L$ with $\delta$, as shown in the second figure (left panel) of Fig.~\ref{fig:paramL_vs_delta}.

We note that all the calculations in the paper are based on the W03 \citep{2003ApJS..144..153W}
fragmentation cross-sections. The impact of using the W03 set or the GAL09 set (provided in the widely
used GALPROP package\footnote{\tiny http://galprop.stanford.edu/web\_galprop/galprop\_home.html}) on the
determination of the halo size $L$ is shown as thin-dotted lines (left panel, same figure)\footnote{The impact
on the transport parameters is detailed in Sect.~7 of \citet{2010A&A...xxx..xxxM}: the region of the best-fit values is slightly displaced, as seen in the figure.}. Any difference existing between these two sets of production cross-sections has no impact on the best-fit value for $L$: thin-dashed
curves (obtained with GAL09 cross-sections) almost match the thick-solid curves (obtained with W03
cross-sections). For other ratios, the effect of the GAL09 cross-sections is always the same,
so it is not discussed further.

\subsubsection{General dependence of $L$ with $\delta$ (for $r_h\neq0$)}
We repeat the analysis with the underdensity $r_h$ as an additional free parameter.
The dependence of $L$ and $r_h$ on the diffusion slope $\delta$ is shown in the right panel
of Fig.~\ref{fig:paramL_vs_delta}. A comparison between the left and the right panel shows
that the combinations of parameters $K_0/L$, $V_a/\sqrt{K_0}$, and $V_c$ are almost unaffected by the
presence of a local bubble; the $\chi^2/$d.o.f is also only slightly affected.

For $\delta\lesssim 0.2$, $r_h$ is consistent with 0 for both model~II (diffusion/reacceleration) and model~III (diffusion/convection/reacceleration). For model~II, the size of $r_h$ suddenly jumps to $\sim 100$~pc. But
for $\delta\gtrsim 0.3$, it returns to the pure diffusion regime, $r_h$ decreasing abruptly (to a non-vanishing value) and $L$ becoming vanishingly small. In this regime, the thin-disc approximation
is no longer valid and nothing can be said about it.
For model~III, the plateau $r_h\sim 100$~pc is stable for all $\delta\gtrsim0.2$. The underdense
bubble also stabilises the value of the halo size $L$. The way of understanding this trend is
as for the standard DM, but now the flux of the radioactive species reads $N_{r_h}^{\rm rad}(0)\propto \exp(-r_h/\sqrt{K\gamma\tau_0})/\sqrt{K\gamma\tau_0}$. The weaker dependence of $L$ with $\delta$
must be represented by this formula. We underline that for all best-fit configurations leading
to $r_h\neq 0$, the improvement is statistically meaningful compared to the case $r_h=0$.

             %%#######################################%%
 \subsection{Isotopic versus elemental measurements}

A similar analysis can be carried out using elemental ratios instead of isotopic ones.
As before, the best-fit values of well-chosen combinations of the transport parameters
$\{K_0,\,\delta,\,V_c,\,V_a\}$
are left unchanged when radioactive species are added to the fit (same values as in
Fig.~\ref{fig:paramL_vs_delta}).

 \subsubsection{General dependence of $L$ with $\delta$}
\begin{figure}[t]
\centering
\includegraphics[width = \columnwidth]{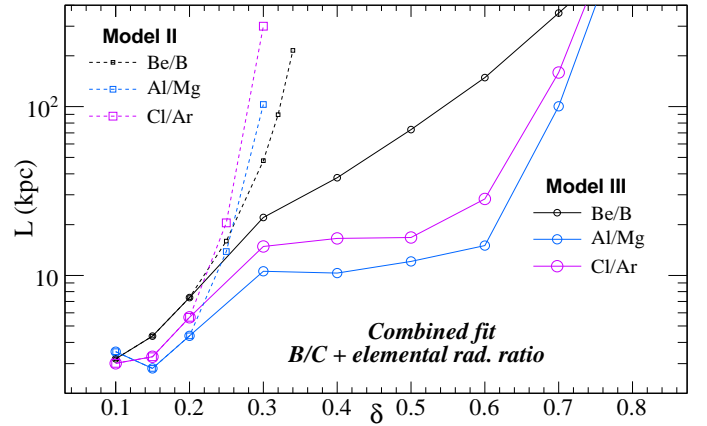}
\caption{Best-fit value of the halo size $L$ as a function of $\delta$ in standard DM, based
on a fit on B/C plus a ratio where a radioactive species is present: B/C+Be/B (black
small symbols), B/C+Al/Mg (blue medium-size symbols), and B/C+Cl/Ar (pink large symbols).
The dashed lines (square symbols) refer to model~II, and the solid lines (circles) refer to model~III.}
\label{fig:paramL_vs_delta2}
\end{figure}
For the standard DM ($r_h=0$), the dependence of the diffusive halo size $L$ on the diffusion slope
$\delta$ is shown in Fig.~\ref{fig:paramL_vs_delta2}, for the three combinations  B/C +
Be/B,  B/C + Al/Mg, and B/C + Cl/Ar.
The trend is similar to that for isotopic ratios:  $L$ increases with increasing $\delta$.
The main difference is that the increase is sharper for both models~II and III. For the former, only
a small region around $\delta\approx 0.2$ corresponds to small halo sizes. For the latter, the halo size
increases sharply above $\delta\gtrsim 0.6$. 

For completeness, similar fits were carried out for the modified DM ($r_h\neq0$). However, adding an
additional degree of freedom only worsens the situation, and the models converge to arbitrarily
small or high values of $L$ and $r_h$. 
Finally, if we fit the combined B/C data, the three isotopic ratios
and the three elemental ratios, we do not obtain more constraints than when fitting B/C and the three
isotopic ratios. This may indicate that the models have difficulties in fitting all these data together:
either the model is incomplete or the data themselves may show some inconsistencies.
This is more clearly seen from the comparison of the model calculation and the data for these elemental
ratios (see below).

\subsubsection{Envelopes of 68\% CL}
\begin{figure}[!t]
	\includegraphics[width = \columnwidth]{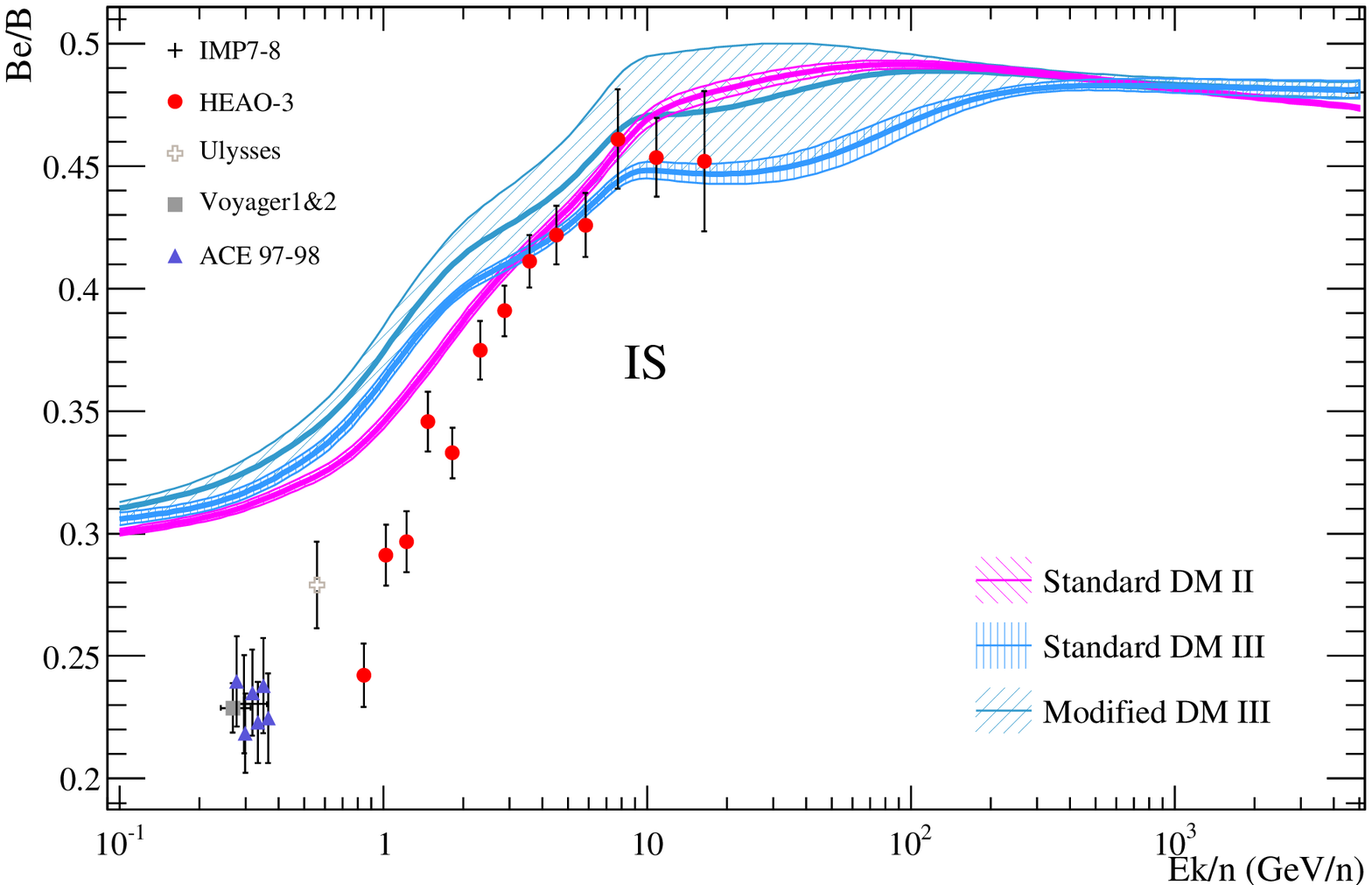}
	\includegraphics[width = \columnwidth]{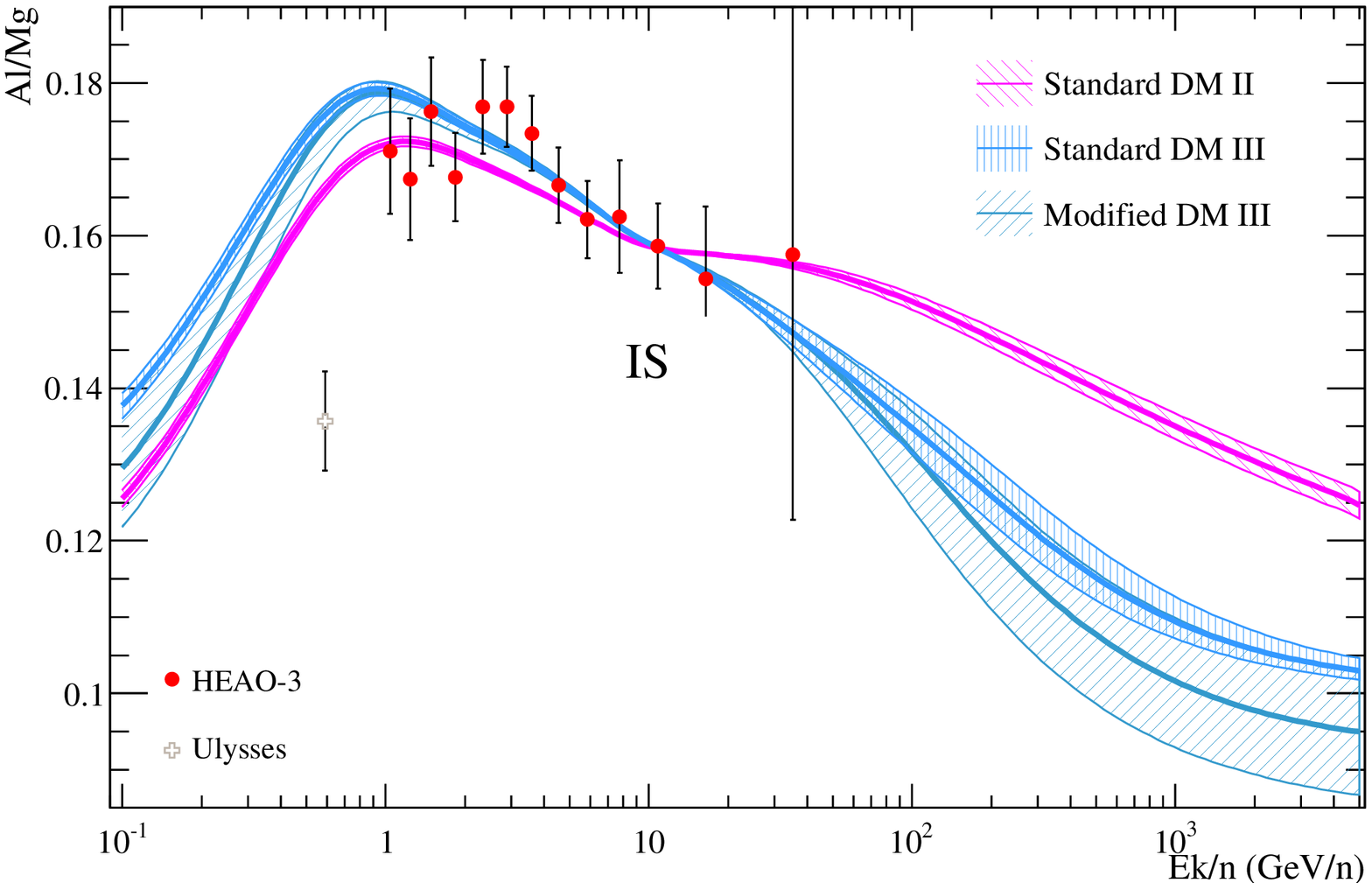}
	\includegraphics[width = \columnwidth]{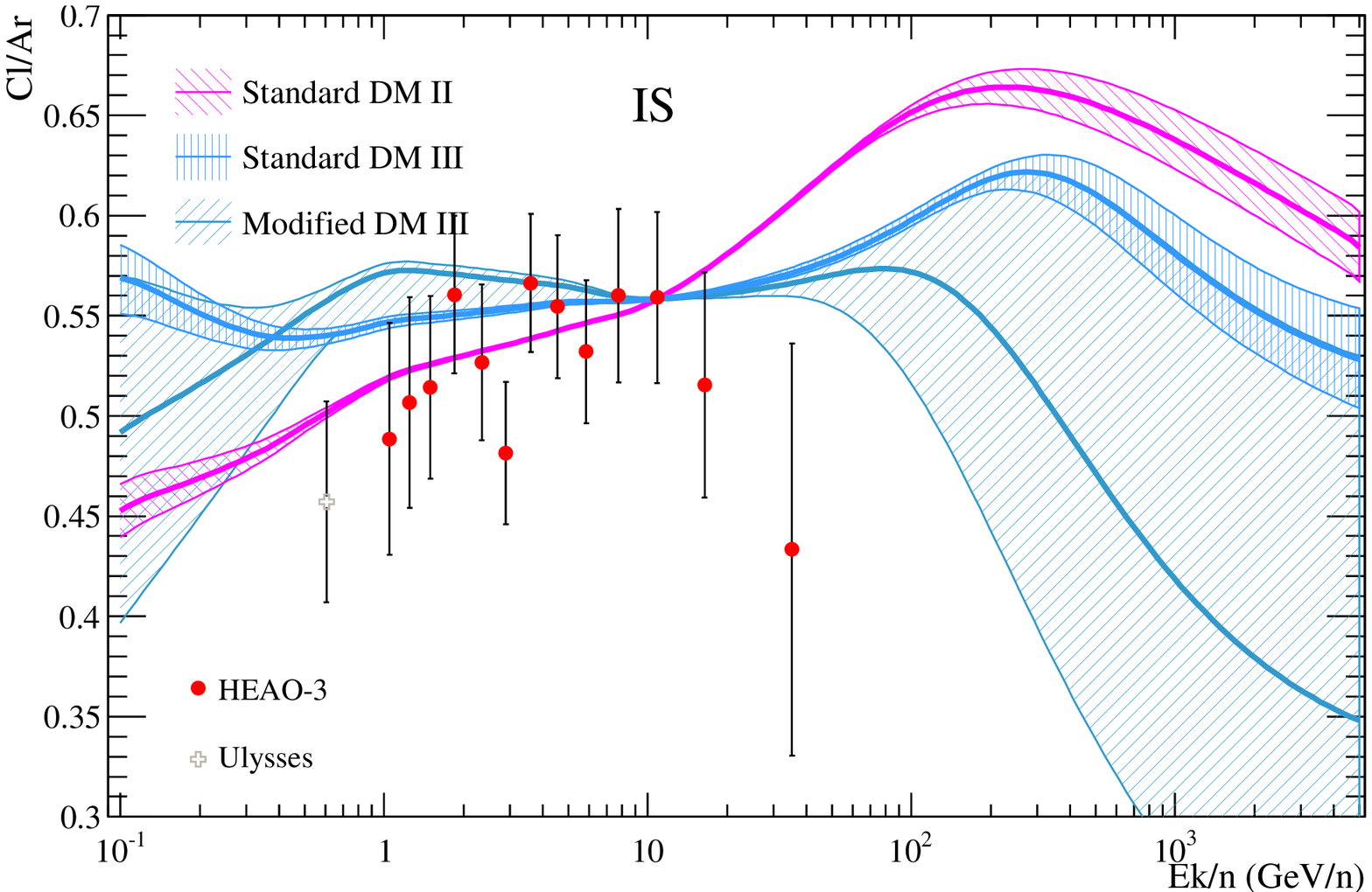}
   \caption{Same as in Fig.~\ref{fig:envelopes} but for the ratios Be/B, Al/Mg, and Cl/Ar.}
	\label{fig:envelopes2}
\end{figure}

From the same set of constraint as in Sect.~\ref{sec:CL} (i.e., B/C and the isotopic ratios of
radioactive species only), we draw the CL for the elemental ratios in Fig.~\ref{fig:envelopes2}.

Given their large error bars, the elemental ratios are in overall agreement with the data,
except at low energy and especially for the Be/B ratio. The main difference between the Be/B ratio and
the two other ratios is that Be and B are pure secondary species, whereas all other elements may contain some primary contribution which can be adjusted to more closely match the data. This also explains why
the Be/B ratio reaches an asymptotical value at high energy (related to the respective production
cross-sections of Be and B), whereas the two others exhibit more complicated patterns. The
low-energy Be/B ratio is related to either the model or the energy biases in the production
cross-sections for these elements (which is still possible, e.g. \citealt{2003ApJS..144..153W}), or to
systematics in the data. To solve this issue, better data over the whole energy range are required.

             %%#######################################%%
 \subsection{Summary and generalisation to the 2D geometry}

Using radioactive nuclei in the 1D geometry, we found that in model II (diffusion/reacceleration),
$L\sim4$~kpc and $r_h\sim 0$, and for the best-fit model III (diffusion/convection/reacceleration),
$L\sim8$~kpc and $r_h\sim120$~pc.  The halo size is an increasing function of the diffusion slope
$\delta$, but in model III the best-fit value for $r_h$ remains $\sim100$~pc for any $\delta\gtrsim
0.3$. This value agrees with direct observation of the LISM (see Appendix~\ref{sec:localbubble}).
Measurement of elemental ratios of radioactive species are not yet precise enough to provide valuable
constraints. 

For now, there are too large uncertainties and too many inconsistencies between the data themselves to enable us to point
unambiguously toward a given model. Moreover, one has to keep in mind that any best-fit model is 
relative to a given set of data chosen for the fit (see Sect.~\ref{sec:sensData}). We note that there
may be ways out of reconciling the low-energy calculation of the Be/B ratio with present data, e.g.,
by changing the low-energy form of the diffusion coefficient \citep{2010A&A...xxx..xxxM}, but this
goes beyond the goal of this paper.

All these trends are found for the models with 2D geometry. We calculate in Table~\ref{tab:1Dvs2D_rad} the
best-fit parameters for the standard model II ($r_h=0$) and the modified model III ($r_h\neq0$). The
values for the 1D geometry are also reported for the sake of comparison. Apart from a few tens of percent
difference in some parameters, as emphasised in Sect.~\ref{sec:summary_stable}, some differences are
expected if the size of the diffusive halo $L$ is larger than the distance to the side boundary
$R$, which is $d_R=12$~kpc in the 2D geometry. It is a well-known result that the closest boundary
limits the effective diffusion region from where CR can originate \citep{2003A&A...402..971T}. For
model~II, $L$ is smaller than $d_R$. We obtain a smaller than 10\% difference for $K_0$, and a $\sim
30\%$ difference for $L$. For the modified model~III, the halo size has a larger scatter (see previous sections),
with $L_{1D}^{\rm best}=13.6>d_R$. The geometry is thus expected to affect the determination
of $L$. We find that $L_{2D}^{\rm best}=4.3$ and that the value of $K_0$ is thus $L_{1D}^{\rm
best}/L_{2D}^{\rm best}\sim 3$ times larger, and $V_a$ is $\sim \sqrt{3}$ times larger than in 1D.

\begin{table}[!tb]
\caption{Best-fit parameters on B/C+$^{10}$Be/$^9$Be$^{26}$+Al/$^{27}$Al +$^{36}$Cl/Cl: 1D vs 2D DM.}
\label{tab:1Dvs2D_rad}\centering
\begin{tabular}{lccccccc} \hline\hline
 Config.     & $\!\!\!\!K_{0} \times 10^2\!\!\!\!$   &  $\delta$  &    $V_{c}$  &  $V_a$ & $L$ & $r_h$ & $\chi^2/$d.o.f\\  
        & $\!\!\!\!\!\!\!\!\!(\unit[]{kpc^2Myr^{-1}})\!\!\!\!\!\!\!\!\!$  &    &
        \multicolumn{2}{c}{$(\unit[]{km~s^{-1})}$}  & \multicolumn{2}{c}{$(\unit[]{kpc})$}\\\hline
& \multicolumn{4}{c}{} \\ 
$\!\!\!$1D-II-$L\!\!\!$          & 15.4 & 0.23 & $\cdots$ & $\!\!$92.5& 6.2   & $\cdots$ & 3.09 \vspace{0.05cm}\\
$\!\!\!$2D-II-$L\!\!\!$          & 14.9 & 0.24 & $\cdots$ & $\!\!$90.8& 8.8   & $\cdots$ & 3.04 \vspace{0.15cm}\\
$\!\!\!$1D-III-$Lr_h\!\!\!\!\!\!$& 1.90 & 0.83 & 18.9$\!\!$ & $\!\!$73.5& 13.6$\!\!$  & $\!\!\!\!0.13\!\!$ & 1.43 \vspace{0.05cm}\\
$\!\!\!$2D-III-$Lr_h\!\!\!\!\!\!$& 5.24 & 0.85 & 18.3  & $\!\!$123.& 4.3$\!\!$  & $\!\!\!\!0.16\!\!$& 1.48 \vspace{0.05cm}\\
\hline
\end{tabular}
\end{table}

%/////////////////////////////////////////////////////////////////
\section{Conclusions\label{sec:conclusion}}
We have used a Markov Chain Monte Carlo technique to extract the posterior distribution functions of
the free parameters of a propagation model. Taking advantage of its sound statistical
properties, we have derived the confidence intervals (as well as confidence contours) of the models for fluxes and other quantities derived from the propagation parameters.

In the first paper of this series (Paper~I), we focused on the phenomenologically
well-understood LBM to ease the implementation of the MCMC. In contrast, here we have analysed a more
realistic DM. In agreement with previous studies, when B/C only is considered, we have confirmed that a
model with diffusion/convection/reacceleration is more likely than the diffusion/reacceleration case.
The former would imply that $\delta\sim 0.8$, whereas the latter would imply that $\delta\sim 0.2$.
This result does not depend on the halo size: we provided simple parameterisations to obtain the
value of the transport parameters for any halo size $L$. If mere eye inspection of the
published AMS-01 data shows consistency with the HEAO-3 data (covering the same energy
region), a B/C analysis based on AMS-01 data (instead of HEAO-3) also indicates that
convection and reacceleration is required, but now providing a diffusion slope
$\delta\sim 0.5$, closer to theoretical expectations. Data from PAMELA or high-energy data from CREAM and TRACER are required to help solving the long-standing uncertainty in the value for $\delta$.

A second important topic of this paper has been the halo size $L$ of the Galaxy and
the impact of the underdense medium in the solar neighbourhood. The determination of
$L$ is for instance crucial to predictions of antimatter fluxes from dark matter 
annihilations. The size of the local underdense medium is as important, as it can bias
the determination of $L$. We provided a step-by-step study of the various radioactive
clocks at our disposal.
Our detailed approach can serve as a guideline as how to take advantage
of future high-precision measurements that will soon become available (e.g., from AMS).
The main conclusions about the constraints provided by the radioactive species are,
in diffusion/reacceleration models, $L\sim4$~kpc and no underdense local bubble
is necessary to match the data. For the best-fit model, which requires diffusion/convection/reacceleration, $L\sim8$~kpc with $r_h\sim120$~pc. For both models, the halo size found is an increasing function of the diffusion slope $\delta$. A striking feature is that in models with convection,
the best-fit value for $r_h$ remains $\sim100$~pc for any $\delta\gtrsim 0.3$. For instance, the B/C
AMS-01 data (which implies that $\delta\sim 0.5$) and the radioactive ratios are consistent with a wind and a local
underdense bubble. This very value of $r_h\sim100$~pc is also supported by direct
observation of the LISM (see Appendix~\ref{sec:localbubble}).

As emphasised in this study, the determination of the value of $L$ and $r_h$ strongly depends
on the value of $\delta$. For all these parameters, high-energy data of 
secondary-to-primary ratios, data in the $\sim$~1~GeV/n$-10$~GeV/n range for isotopic ratios
(of radioactive species), and/or data for the radioactive elemental ratios in the
$1-100$~GeV/n energy range are necessary. This is within reach of several flying and forthcoming
balloon-borne projects and satellites (PAMELA, AMS).

%****************************************************************************
%****************************************************************************
%****************************************************************************
\begin{acknowledgements}
We thank C. Combet for a careful reading of the paper.
A.~P is grateful for financial support from the Swedish Research
Council (VR) through the Oskar Klein Centre.
We acknowledge the support of the French ANR (grant ANR-06-CREAM).
\end{acknowledgements}

%****************************************************************************
%****************************************************************************
%\onecolumn
\begin{appendix}

%****************************************************************************
\section{Solutions of the diffusion equation\label{App:solutions}}

We provide below the solutions for the diffusion equation with a constant wind $V_c$
and a single diffusion coefficient $K(E)$ in the whole Galaxy.
In the 1D version of the model (e.g., \citealt{2001ApJ...547..264J}),
the source distribution and the gas density do not depend on $r$, so
that the propagated fluxes depend only on $z$.

The derivation of these solutions is very similar and has no additional
difficulties to those experienced by~\citet{2001ApJ...555..585M},
to which we refer the reader for more details. As
both frameworks (1D and 2D) exhibit similar forms, formulae are written for
the 1D model only. Formulae for the 2D case  are obtained by replacing some 1D
quantities by their 2D counterparts, as specified below.

               %%%%%%%%%%%%%%%%%%%%%%%%%%%

\subsection{1D--model}

The starting point is the transport
equation \citep{1990acr..book.....B}. We assume that the
diffusion coefficient $K$ does not depend on spatial coordinates.
A constant wind $V_c$ blows the particle away from the Galactic disc,
along the $z$ direction. In the thin-disc approximation (e.g., \citealt{1992ApJ...390...96W}),
the diffusion/convection for the 1D--model (discarding energy redistributions)
is
\begin{equation}
\label{1D-exotic-thin-disc}
  \displaystyle \left\{  -K \frac{d^2}{dz^2}
 +V_c \frac{\partial }{\partial z} + \Gamma_{\rm rad} + 2h\Gamma_{\rm tot}\delta(z)\right\} N 
 \equiv {\cal L} N = {\cal Q}(z).
\end{equation}
In this equation, $N$ is the differential density of a given CR species,
$\Gamma_{\rm rad}=1/(\gamma\tau)$ is its decay rate, and
$\Gamma_{\rm tot}= \sum_{\rm ISM} n_{\rm ISM} v\sigma_{\rm ISM}$
is its destruction rate in the thin gaseous disc ($n_{\rm ISM}=$H, He). 
The right-hand side (r.h.s) of the equation is a generic source term, that contains one of the
following three contributions, i.e., ${\cal Q}(z)={\cal P}(z)+{\cal S}(z)+{\cal R}(z)$:
  \begin{description}
	  \item[i) ${\cal P}(z)=2h\delta(z)\times q^s_0 Q(E) $] is the standard primary source term for sources
		located in the thin disc. The quantity $q^s_0$ is
		the source abundance of nucleus $j$ whose source spectrum is $Q(E)\propto \beta^{\eta} {\cal R}^{-\alpha}$.
		\item[ii) ${\cal S}(z)=2h\delta(z)\times \Gamma_{\rm tot}^{p\rightarrow s}N^p(z=0,E)$] is the standard secondary source
		term (also in the disc), 
		where $\Gamma_{\rm tot}^{p\rightarrow s}=n v\sigma^{p\rightarrow s}$ is the production cross-section of nucleus $p$
		into $s$. This simple form originates from the straight-ahead approximation used when dealing with
		nuclei (see, e.g., ~\citealt{2001ApJ...555..585M} for more details). 
		\item[iii) ${\cal R}(z) = \Gamma_{\rm decay}^{r\rightarrow s'}N^r(z,E)$] described
    a contribution from a radioactive nucleus $r$, decaying into $s'$ in both the disc and the halo. 
	\end{description}
The equation is even in $z$ so that it is enough to solve it in the upper-half plane.
The use of the standard boundary condition $N(z=L)=0$ and continuity of the density and the current
at the disc crossing completely characterises the solution.

                            %-----------------%
	\subsubsection{Stable species}
For a mixed species, primary and secondary standard sources add up, so that,
for a nucleus $k$ with no radioactive contribution, the source term is rewritten as
\begin{equation}
\label{eq:Q}
  Q^m_{\rm disc}(E)= q^m_0 Q(E) + \sum_{k>m}\Gamma_{\rm tot}^{k\rightarrow m}N^p(z=0,E),
\end{equation}
and the corresponding equation to solve is then
\[
 {\cal L}^m N^m= 2h\delta(z)\cdot Q^m_{\rm disc}(E)\;.
\]
We find the solution in the halo, apply the boundary
condition $N(z=L)=0$, and then ensure continuity between
the disc and the halo, so that
\begin{equation}
\label{eq:refNz}
N^m(z)=  N^m(0) \cdot \exp^{(V_cz/2K)}  \frac{\sinh(S^m(L-z)/2)}{\sinh(S^mL/2)}
\end{equation}
and
\begin{equation}
\label{eq:refN0}
   N^m(0)= \frac{2h Q^m_{\rm disc}(E)}{A^m}\;.
\end{equation}
The quantities $S^m$ and $A^m$ are defined as 
\begin{eqnarray}
S^m &\equiv& \sqrt{\frac{V_c^2}{K^2} + 4 \frac{\Gamma_{\rm rad}^m}{K^2}}\;;\\
A^m &\equiv& V_c \!+\! 2h\Gamma^m_{\rm tot} \!+\! K S^m\!\coth\!\left(\!\frac{S^mL}{2}\!\right)\!.
\end{eqnarray}

                            %-----------------%
	\subsubsection{Adding a $\beta$-decay source term: general solution}
It is emphasised in \citet{2001ApJ...555..585M} that the $^{10}$Be$\rightarrow$$^{10}$B channel
contributes up to 10\% in the secondary boron flux at low energy and cannot
be neglected. Although the spatial distribution of a radioactive nucleus decreases exponentially
with $z$, we have to consider that the source term is emitted from the halo,
complicating the solution. The equation to solve for the nucleus $j$, which is
$\beta$-fed by its radioactive parent $r$ is
\[
 \displaystyle {\cal L}^j N^j =  \Gamma_{\rm decay}^{r\rightarrow j}N^r(z,E)\;,
\]
where $N^r(z,E)$ is given by Eq.~(\ref{eq:refNz}).
The solution is found following the same steps as above, although
it has a more complicated form (due to a non-vanishing source term
in the halo).

If we take into account both the standard source term $Q^j_{\rm disc}(E)$
and the radioactive contribution of the nucleus $N^r$, we obtain:
\begin{eqnarray}
N^j(z)&=&  \left\{  \varpi \cdot \frac{\sinh(S^j(L-z)/2)}{\sinh(S^jL/2)} \right.\nonumber \\ 
 &&\left. -\nu \Theta \Lambda \cdot \frac{\cosh(S^jz/2)}{\cosh(S^jL/2)}\right\} \times \exp^{(V_cz/2K^j)} \nonumber \\ 
+ &\Theta& \left\{ \lambda \sinh\left(\frac{S^r}{2}(L-z)\right) +\Lambda \cosh \left(\frac{S^r}{2}(L-z)\right) \right\} \nonumber \\
&& ~~~~~~~~~~~~~~~~~~~~~~~~~~~~~~\times \exp^{(V_cz/2K^r)}, \label{eq:Ngen}
\end{eqnarray}
where
\begin{equation}
\Theta \equiv  -\frac{\Gamma^{k\rightarrow j}_{\rm rad}}{K^j(\lambda^2-\Lambda^2)} \frac{N^r(0)}{\sinh(S^rL/2)}
\end{equation}
and
\begin{eqnarray}
\label{eq:mu}
 \varpi &\equiv& \frac{2h Q^j_{\rm disc}}{A^j} +  \frac{\Theta}{A^j} \times \\
	&&\left\{ \frac{\nu a_i}{\cosh(S^jL/2)}\left[V_c + 2h\Gamma^j \right]  \right.\nonumber\\
  && -\sinh(S^rL/2)\left[a\left( V_c(2-\frac{K^j}{K^r}) + 2h\Gamma^j\right) +a_i  K^jS^r\right] \nonumber\\
 -  &&\left.	\cosh(S^rL/2)\left[a_i\left( V_c(2-\frac{K^j}{K^r})  + 2h\Gamma^j\right) +a K^jS^r\right] \right\}, \nonumber
\end{eqnarray}
where
\begin{equation}
\kappa \equiv 1/K^r-1/K^j\nonumber
\end{equation}
\begin{equation}
\nu\equiv e^{\kappa V_cL/2}
   \quad
\Lambda\equiv \kappa\frac{S^rV_c}{2}
   \quad 
\lambda \equiv \frac{\kappa V_c^2}{2K^r}  + \frac{\Gamma_{\rm rad}^r}{K^r} - \frac{\Gamma_{\rm rad}^j}{K^j}.\nonumber
\end{equation}
The superscript on $K$ indicates that the diffusion coefficient is to be
evaluated at a rigidity calculated for the nucleus $m$. The latter can differ
from one nucleus to another because, the calculation is performed
at the same kinetic energy per nucleon for all the nuclei (hence at slightly
different rigidities for different nuclei). 
To compare with the data, the flux is calculated at $z=0$
\begin{equation}
\label{eq:N0rad}
N^j(0) \!=\! \varpi + \Theta \left[\! \lambda\sinh (\frac{S^rL}{2}) \!+\! \Lambda \cosh (\frac{S^rL}{2})
\!-\! \frac{\nu \Lambda}{\cosh \frac{S^jL}{2}} \!\right].
\end{equation}

                           %-----------------%
	\subsubsection{Solution including energy redistributions}
When energy redistributions are included, the solution $N^h(z)$ in the halo remains
the same because our model assumes no energy redistributions in that region.
Only the last step of the calculation changes (ensuring continuity during the disc crossing).
The new solution is denoted ${\cal N}(0)$.

For the case of a mixed species $m$ without radioactive contribution, the result
is straightforward: the solution for the halo is still given by Eq.~(\ref{eq:refNz}),
but ${\cal N}^m(0)$ is now given by
\begin{equation}
 {\cal N}^m(0) \!=\! N^m(0) - \frac{2h}{A^m}\left( b(E) \frac{d{\cal N}^m(0)}{dE} + c(E) \frac{d^2{\cal N}^m(0)}{dE^2}\right)\,,\nonumber
\end{equation}
which is solved numerically, $N^m(0)$ being the solution
when energy terms are discarded, i.e., Eq.~(\ref{eq:refN0}).
The terms $a(E)$ and $b(E)$ describing energy losses and gains are
discussed in Sect.~\ref{s:transport}.

When a radioactive contribution exists, the constant left to determine
is $\varpi$ from Eq.~(\ref{eq:Ngen}), which we denote now $\varpi^*$
\begin{equation}
\label{eq:varpistar}
 \varpi^* = \varpi - \frac{2h}{A^j}\left( b(E) \frac{d{\cal N}^j(0)}{dE} + c(E) \frac{d^2{\cal N}^j(0)}{dE^2}\right)\;.
\end{equation}
As above, $\varpi$ denotes the quantity evaluated without energy redistribution,
whereas ${\cal N}^j(0)$ denotes the equilibrium flux at $z=0$.
To ensure ${\cal N}^j(0)$ also appears in the l.h.s. of the equation,
we form the quantity
\begin{equation}
\Xi \equiv \Theta \left[\! \lambda\sinh (\frac{S^rL}{2}) \!+\! \Lambda \cosh (\frac{S^rL}{2})
\!-\! \frac{\nu \Lambda}{\cosh \frac{S^jL}{2}} \!\right].
\end{equation}
Hence $N^m(0) = \varpi + \Xi$, and we can add to both sides of Eq.~(\ref{eq:varpistar})
the quantity $\Xi$, so that we recover the standard form
\[
 {\cal N}^j(0) = N^j(0) - \frac{2h}{A^j}\left( b(E) \frac{d{\cal N}^j(0)}{dE} + c(E) \frac{d^2{\cal N}^j(0)}{dE^2}\right)\;,
\]
which we solve numerically.

This is the solution in the disc ($z=0$). The solution for any $z$
is obtained from Eq.~(\ref{eq:Ngen}), making the substitution
\begin{equation}
	\varpi \rightarrow \varpi^* = {\cal N}^j(0) - \Xi.
\end{equation}
We note that for $\Theta=0$ (i.e., no radioactive contribution)
the result for standard sources in the disc is recovered.

               %%%%%%%%%%%%%%%%%%%%%%%%%%%

\subsection{2D geometry}
\label{app:2D}
Cylindrical symmetry is now assumed, both the CR density $N$ and the source
terms depending on $r$. Compared to Eq.~(\ref{1D-exotic-thin-disc}),
the operator $\triangle_r$ now acts on $N(r,z)$. 

An expansion along the first order Bessel function is performed
\begin{equation}
\label{eq:Niresum}
N(r,z) = \sum_{i=1}^{\infty} N_i(z) J_0\left(\zeta_i \frac{r}{R}\right)\;.
\end{equation}
The quantity $\zeta_i$ is the $i$-th zero of $J_0$, and this form automatically
ensures the boundary condition $N(r=R,z)=0$. We have
\[
-\triangle_r J_0\left(\zeta_i \frac{r}{R}\right) = \frac{\zeta_i}{R^2} J_0\left(\zeta_i \frac{r}{R}\right),
\]
so that each Bessel coefficient $N_i(z)$ follows an equation very similar to
Eq.~(\ref{1D-exotic-thin-disc}), where 
\[
	\Gamma_{\rm rad} \Rightarrow \Gamma_{\rm rad} + \frac{\zeta_i}{R^2},
\]
and where each source term must also be expanded on the Bessel basis.
More details can be found in \citet{2001ApJ...555..585M}.

The full solutions for mixed species, with stable or radioactive
parents, is straightforwardly obtained from 1D ones, after making
the substitutions
\begin{equation}
  N^j(z) \stackrel{\rm 2D~model}{\Longrightarrow}  N^j_i(z),
\end{equation}
\begin{equation}
S^j  \stackrel{\rm 2D~model}{\Longrightarrow}
   S^j_i\!\!\equiv\!\!\sqrt{\frac{V_c^2}{K^2} \!+\! 4\frac{\zeta_i^2}{R^2} 
    + 4 \frac{\Gamma_{\rm rad}^j}{K^2}} \;,
\end{equation}
\begin{equation}
A^j  \stackrel{\rm 2D~model}{\Longrightarrow} 
    A^j_i \equiv 2h \Gamma^j + V_c + KS^j_i\coth\left(\frac{S^j_iL}{2}\right),
\end{equation}
and
\begin{equation}
\Theta^j(S^r,N^r(0)) {\Longrightarrow} \Theta_i^j(S_i^r,N^r_i(0))\;,
\end{equation}
\begin{equation}
\varpi^j(S^j,A^j) {\Longrightarrow} \varpi_i^j(S_i^j,A^j_i)\;,
\end{equation}
\begin{equation}
\lambda^r(S^r) {\Longrightarrow} \lambda_i^r(S_i^r)\;.
\end{equation}

The above formulae, for the radioactive source, differ slightly from those
presented in \citet{2001ApJ...555..585M}. However, the only difference is
in the flux for $z\neq 0$, which was not considered in this paper.

               %%%%%%%%%%%%%%%%%%%%%%%%%%%
	\section{The local bubble\label{sec:localbubble}}
The underdensity in the local interstellar matter (LISM) is coined the
local bubble\footnote{For a state-of-the-art view on the subject, the
reader is referred to the proceedings of a conference held in 2008:{\em The Local Bubble and
Beyond II} | http://lbb.gsfc.nasa.gov/}.
The LISM is a region of extremely hot gas ($\sim
10^5-10^6$ K) and low density ($n\lesssim$ 0.005 cm$^{-3}$) within an asymmetric bubble of
radius $\lesssim$ 65-250 pc surrounded by dense neutral hydrogen walls
\citep{1999A&A...346..785S,2000ApJ...528..756L,2000ApJ...534..825R}. This picture has been
refined by subsequent studies, e.g., \citet{2003A&A...411..447L}. 
The Sun is located inside a local interstellar cloud (LIC) of typical extension $\sim 50$ pc
whose density $N_{HI}\sim$ 0.1 cm$^{-3}$ \citep{2004A&A...426..845G,2008ApJ...683..207R}.
Despite these successes, a complete mapping and understanding of the position and properties
of the gas/cloudlets filling the LISM, as well as the issue of interfaces with other bubbles
remains challenging (e.g., \citealt{2008ApJ...673..283R,2008A&A...486..471R}).
Based on existing data, numerical simulations of the local bubble infer that it is the result of $14-19$
SNe occurring in a moving group, which passed through the present day local H$_{\rm I}$ cavity
$13.5-14.5$~Myr ago \citep{2006A&A...452L...1B}. The same study suggests that the local bubble expanded
into the Milky Way halo roughly 5 Myr ago.

A last important point, is that of the existence of turbulence in the LISM to scatter off CRs.
The impact of the underdense local bubble on the production of radioactive nuclei as modelled in
Eq.~(\ref{eq:rad_damping}) depends whether the transport of the radioactive nuclei in this
region is diffusive or not. In a study based on a measurement of the radio
scintillation of a pulsar located within the local bubble, \citet{2008SSRv..tmp...98S} infers that
values for the line of sight component of the magnetic field are only slightly less, or
completely consistent with, lines of sight through the general interstellar medium; the
turbulence is unexpectedly high in this region.

These pieces of observational evidences support the model used in Sect.~\ref{sec:rad},
leading to an enhanced decrease in the flux of radioactive species at low energy. A
detailed study should take into account the exact morphology of the ISM (asymmetry, cloudlets).
However, there are so many uncertainties in this distribution and the associated level
of turbulence, that a crude description is enough to capture a possible effect in the CR data.

               %%%%%%%%%%%%%%%%%%%%%%%%%%%
	\section{MCMC optimisation\label{sec:MCMC_combi}}
The efficiency of the MCMC increases when the PDFs of the parameters are close to resembling Gaussians. Large
tails in PDFs require more steps to be sampled correctly. A usual task in the MCMC machinery is to
find some combinations of parameters that ensure that these tails disappear. This was not discussed in the
case of the LBM as the efficiency of the PDF calculation was satisfactory. In 1D (or 2D) DMs,
the computing time is longer and the efficiency is found to be lower. To optimise and speed up
the calculation, we provide combinations of parameters that correspond to a Gaussian distribution.
\begin{figure*}[!t]
\centering
\includegraphics[width = 0.47\textwidth]{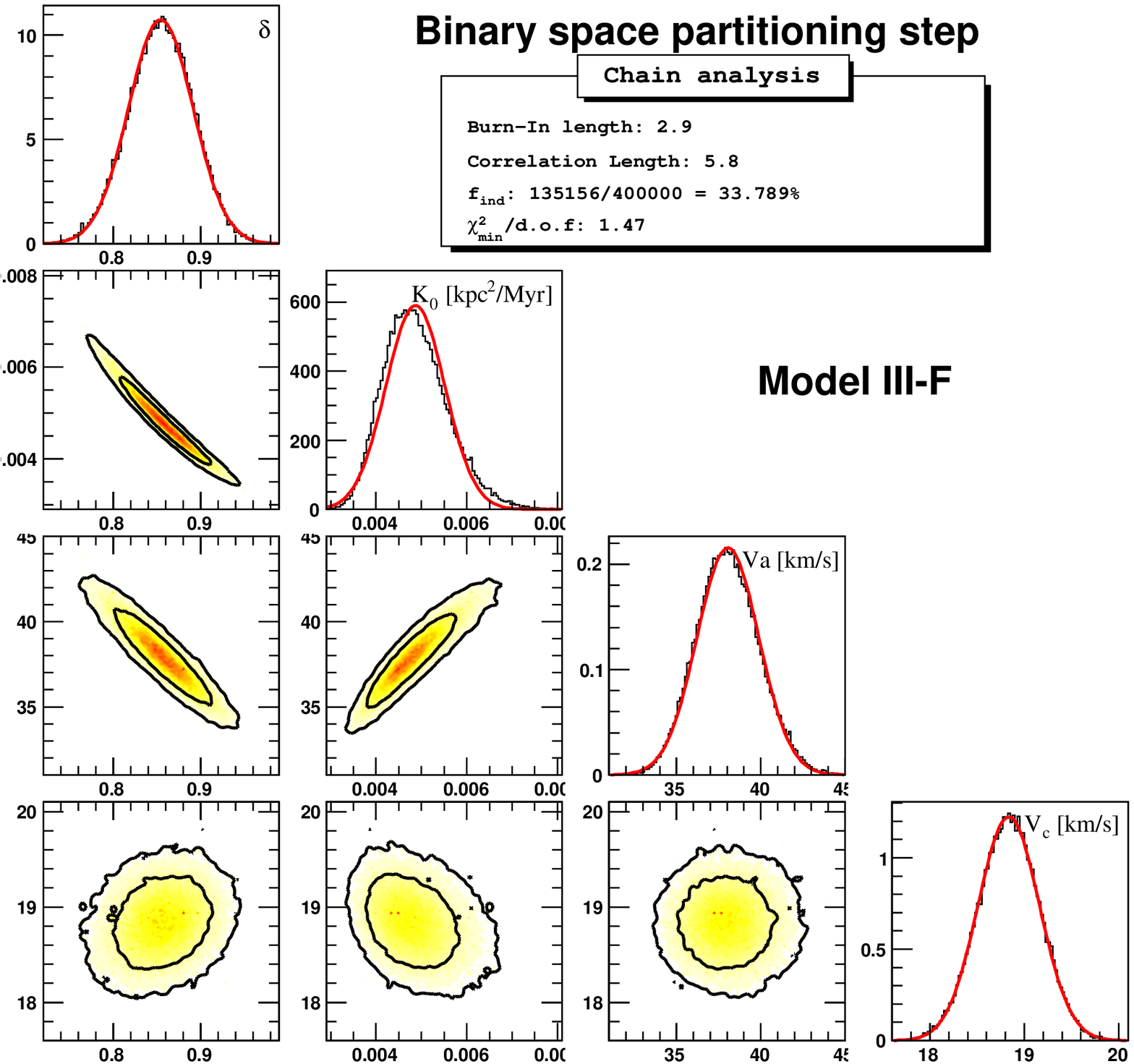}\vspace{3mm}
\includegraphics[width = 0.47\textwidth]{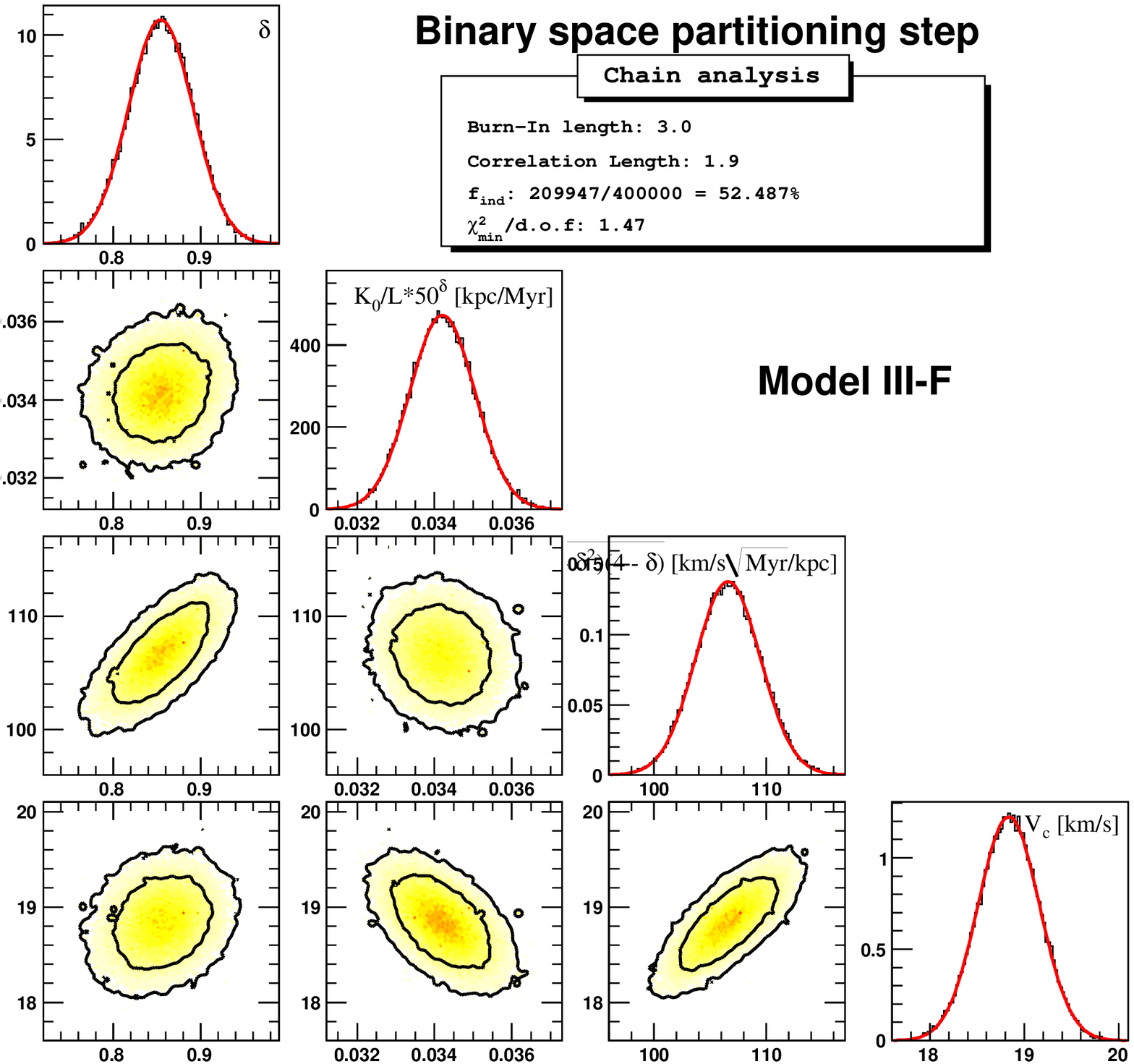}
\caption{Posterior PDFs of the model parameters (using the Binary Space Partitioning step|see Paper~I,
and the B/C constraint).
The diagonals show the 1D marginalised PDF of the indicated parameters, and the red line results from a
Gaussian fit to the histogram. Off-diagonal plots show the 2D marginalised posterior PDFs for the
parameters in the same column and same line, respectively. The colour code corresponds to the regions of
increasing probability (from paler to darker shades), and the two contours (smoothed) delimit regions
containing respectively 68\% and 95\% (inner and outer contour) of the PDF. Left panel: PDFs for
$\{V_c,\,\delta,\,K_0,\,V_a\}$. Right panel: the same PDFs but shown for a different combination of the
parameters $\{V_c,\,\delta,\,K_0/L\times50^\delta,\,V_a/\sqrt{K_0\times3\delta(4 - \delta^{2})(4 -
\delta)}\}$.}
\label{fig:MCMC_combi}
\end{figure*}

A typical PDF determination with four free parameters $\{V_c,\,\delta,\,K_0,\,V_a\}$ (see next
section) is shown in Fig.~\ref{fig:MCMC_combi}.
The diagonal of the left panel shows the PDF of these parameters (black histogram), on which a Gaussian fit is
superimposed (red line). We see a sizeable tail for the $K_0$ parameter, and a small asymmetry for
the $V_a$ parameter. The right panel shows the same PDFs, but for the following combinations of
the transport parameters:
\begin{equation}
  K_0 \quad \longleftrightarrow  \quad \frac{K_0}{L}\times 50^\delta
\label{eq:combi_K0}
\end{equation}
\begin{equation}
  V_a  \quad \longleftrightarrow  \quad \frac{V_a}{\sqrt{K_0\times3\delta(4 - \delta^{2})(4 -\delta)}}
\label{eq:combi_Va}
\end{equation}
These forms are inspired by the known degeneracies between parameters. For instance, in diffusion models,
the secondary to primary ratio is expected to remain unchanged as long as the {\em effective} grammage
$\langle x\rangle$ of the model is left unchanged. For pure diffusion, we obtain 
\citep{2001ApJ...547..264J,2006astro.ph.12714M} for the grammage $\langle x\rangle =\Sigma vL/(2K)=\Sigma
L/(2cK_0{({\cal R}/1{\rm GV})}^\delta)$, where $\Sigma$ is the surface density.  Apart from the $K_0-L$
degeneracy, the parameters $K_0$ and $\delta$ are correlated. We find that the combination
$K_0\times50^\delta$ is appropriate for removing the $K_0$ PDF's tail (see Fig.~\ref{fig:MCMC_combi}). The
origin of the value $50$ is unclear. It may be related to the energy range covered by B/C HEAO-3 data on
which the fits are based.
The combination used for $V_a$ [Eq.~(\ref{eq:combi_Va})] comes directly from the form of the
reacceleration term Eq.~(\ref{eq:Va}). Reacceleration only plays a role at low energy, so we can take
${\cal R}^\delta\approx1$ and end up with the combination presented in Eq.~(\ref{eq:combi_Va}).
The independent acceptance $f_{\text{ind}}$, defined in Paper~I as the ratio of the number of independent 
samples to the total step number, increases from 1/3 to 1/2 by using the above described parameter 
combinations for the four parameter model presented in Fig.~\ref{fig:MCMC_combi}.

A last combination is for the local bubble parameter $r_h$:
\begin{equation}
  r_h  \quad \longleftrightarrow  \quad \frac{r_h}{\sqrt{K_0}}\;.
\label{eq:combi_rh}
\end{equation}
This comes from the form of Eq.~(\ref{eq:rad_damping}),
where the flux damping for radioactive species (due to the local bubble) is
effective only at low energy ($\gamma\approx1$, ${\cal R}\approx 1$).

            %%#######################################%%
  \section{Datasets for CR measurements}

            %%#######################################%%
  \subsection{B/C ratio\label{datasetBC}}
Unless specified otherwise, the reference B/C dataset used throughout the paper is
denoted dataset~F: it consists of i) low-energy data taken by the
IMP7-8~\citep{1987ApJS...64..269G}, the Voyager~1\&2 \citep{1999ICRC....3...41L}, and
the ACE-CRIS \citep{2006AdSpR..38.1558D} spacecrafts; ii) intermediate energies
acquired by HEA0-3 data \citep{1990A&A...233...96E}; and iii) higher energy data from
Spacelab~\citep{1990ApJ...349..625S} and the published CREAM
data~\citep{2008APh....30..133A}.  Other existing data are discarded either because of
their too large error bars, or because of their inconsistency with the above data (see
Paper~I).

            %%#######################################%%
  \subsection{Isotopic and elemental ratios of radioactive species\label{app:data_rad}}
For $^{10}$Be/$^{9}$Be, the data are taken from balloon flights
\citep{1977ApJ...212..262H,1978ApJ...226..355B,1979ICRC....1..389W}, including the 
ISOMAX balloon-borne instrument \citep{2004ApJ...611..892H}, and from the IMP-7/8
\citep{1977ApJ...217..859G}, ISEE-3 \citep{1980ApJ...239L.139W}, Ulysses \citep{1998ApJ...501L..59C},
Voyager \citep{1999ICRC....3...41L}, and ACE spacecrafts \citep{2001ApJ...563..768Y}.
For $^{26}$Al/$^{27}$Al, the data consist of a series of balloon flights \citep{1982ApJ...252..386W},
and the ISEE-3 \citep{1983ICRC....9..147W}, Voyager \citep{1994ApJ...430L..69L}, Ulysses
\citep{1998ApJ...497L..85S}, and ACE spacecrafts \citep{2001ApJ...563..768Y}.
For $^{36}$Cl/Cl, the data are from the CRISIS balloon \citep{1981ApJ...246.1014Y}, and from the
Ulysses \citep{1998ApJ...509L..97C} and ACE \citep{2001ApJ...563..768Y} spacecrafts.

The data for the elemental ratios come from the HEAO-3 \citep{1990A&A...233...96E},
Ulysses \citep{1996ApJ...465..982D}, and the ACE spacecrafts \citet{2006AdSpR..38.1558D}. 
The published ACE data on Al/Mg and Cl/Ar \citep{2009ApJ...698.1666G} were not
used as Be/B is not provided.

\end{appendix}

%%%%%%%%%%%%%%%%%%%%%%%%%%%%%%%%%%%%%%%%%%%%%%%%%%%%%%%%%%%%%%%%%%%%%%%%
%%%%%%%%%%%%%%%%%%%%%%%%%%%%%%%%%%%%%%%%%%%%%%%%%%%%%%%%%%%%%%%%%%%%%%%%
\bibliographystyle{aa}
\bibliography{mcmcII}
\end{document}